\documentclass[aps,twocolumn,floatfix,superscriptaddress,preprintnumbers]{revtex4}
\usepackage{amsfonts}
\usepackage{amssymb}
\usepackage{amsmath,mathtools}
\usepackage{color}
\usepackage{ulem}

\RequirePackage{ifpdf}
\ifpdf
  \usepackage[pdftex]{graphicx}
\else
  \usepackage[dvipdfmx]{graphicx}
\fi

\newcommand{\braket}[1]{{\langle #1 \rangle}}

\usepackage{slashed}
\usepackage{simplewick}
\DeclareMathOperator{\tr}{tr}

\begin{document}

\title{Deep Learning and Holographic QCD}
\author{Koji Hashimoto}
\author{Sotaro Sugishita}
\affiliation{Department of Physics, Osaka University, Toyonaka, Osaka 560-0043, Japan}
\author{Akinori Tanaka}
\affiliation{Mathematical Science Team, RIKEN Center for Advanced Intelligence Project (AIP),1-4-1 Nihonbashi, Chuo-ku, Tokyo 103-0027, Japan}
\affiliation{Department of Mathematics, Faculty of Science and Technology, Keio University, 3-14-1 Hiyoshi, Kouhoku-ku, Yokohama 223-8522, Japan}
\affiliation{interdisciplinary Theoretical \& Mathematical Sciences Program (iTHEMS) RIKEN 2-1, Hirosawa, Wako, Saitama 351-0198, Japan}
\author{Akio Tomiya}
\affiliation{Key Laboratory of Quark \& Lepton Physics (MOE) and Institute of Particle Physics,
Central China Normal University, Wuhan 430079, China}
\affiliation{RIKEN/BNL Research center, Brookhaven National Laboratory, 
Upton, NY, 11973, USA}

\begin{abstract}
We apply 
the relation between deep learning (DL) and the AdS/CFT correspondence to a holographic model of QCD.
Using a lattice QCD data of the chiral condensate at a finite temperature as our training data,
the deep learning procedure holographically determines an emergent bulk metric as neural network weights. 
The emergent bulk metric is found to have both a black hole horizon and a finite-height IR wall,
so shares both the confining and deconfining phases, signaling the cross-over thermal phase transition
of QCD.
In fact, a quark antiquark potential holographically calculated by the emergent bulk metric
turns out to possess both the linear confining part and the Debye screening part,
as is often observed in lattice QCD. From this we argue the discrepancy
between the chiral symmetry breaking and the quark confinement in the holographic QCD.
The DL method is shown to provide a novel data-driven holographic modeling of QCD, and sheds light on
the mechanism of emergence of the bulk geometries in the AdS/CFT correspondence.

\end{abstract}


\maketitle

\setcounter{footnote}{0}

\noindent
\section{Introduction.}
\label{sec:1}

Holographic modeling of QCD has provided novel ways to look at various 
hadronic phenomena of QCD, and has constructed a firmer bridge between
strongly coupled quantum field theories and classical/quantum gravity.
The bottom-up approach of holographic QCD, initiated in Refs.~\cite{Erlich:2005qh,DaRold:2005mxj},
uses the dictionary of the renowned AdS/CFT correspondence
\cite{Maldacena:1997re,Gubser:1998bc,Witten:1998qj} to construct a 5-dimensional
gravity model of QCD. Quite simple holographic models, together with various further 
refinement, capture nicely non-perturbative properties of QCD, ranging from hadron spectra to
condensates, inter-quark forces, phase transitions and non-equilibrium dynamics.

Generic obstacles in any model building is to solve inverse problems.
Phenomenological models, based on specific Hamiltonians, Lagrangians or equations,
include many parameters which are determined once the calculated results of physical 
observables are compared with experiments. Therefore, if the number of
parameters is large (and in general it could be infinite), determining model parameters
is increasingly difficult. Holographic QCD models are of course of this sort --- 
even worse, as a bulk gravity metric is necessary to define a model, and the metric has a functional
degree of freedom. This is the inverse problem: From given experimental data, how 
can we fix the bulk gravity metric?

Conventional holographic QCD models assume a gravity metric in the 5-dimensional bulk
spacetime. Typically, to describe the quark gluon plasma phase at a high temperature,
one uses AdS-Schwarzschild black hole metric. On the other hand, the confining 
phase at a low temperature is described by so-called confining geometries, 
which end with an IR wall. At any case, one should assume a metric to define a
holographic model, then calculate physical observables using the AdS/CFT dictionary,
and compare them with experimental values.

\begin{figure}
\includegraphics[width=8.5cm]{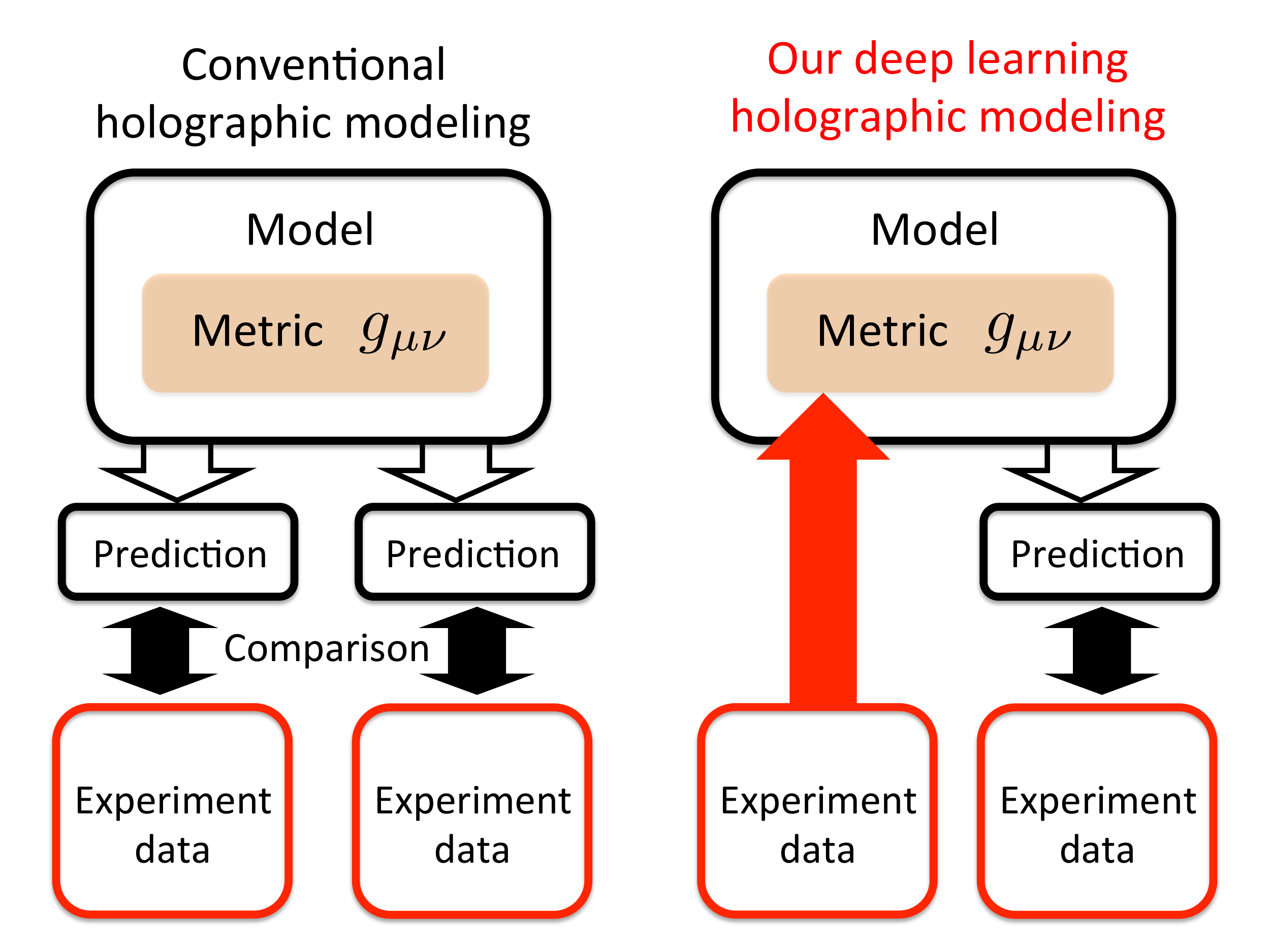}
\caption{A schematic view of our deep learning holographic modeling, emphasizing the difference
from the conventional holographic modeling.}
\label{fig:model}
\end{figure}

We here employ the virtue of the deep learning (DL) \cite{Hinton,Bengio,LeCun}. 
In our previous paper \cite{Hashimoto:2018ftp},
we provide a mapping between the AdS/CFT and a deep neural network, and identify 
the bulk metric with the weights of the network, to generate an emergent metric from
given data of the boundary quantum field theory. This AdS/DL correspondence is suitable
for holographic QCD modeling, as it can solve the inverse problem. Fig.~\ref{fig:model}
describes schematically the difference between the conventional and our deep-learning
holographic modelings. We first use a certain QCD data to determine the bulk metric (and also some other parameters of the model) via deep learning, and then calculate other physical
observables of QCD by the model with the determined metric, as a prediction of the model.

Specifically, we study a $\phi^4$ theory in the bulk, to describe the chiral condensate of QCD.
We use lattice QCD data \cite{Unger} of the chiral condensate as a function of the quark mass,
at a finite temperature near the critical temperature of the thermal phase transition.
The DL determines the bulk metric which reproduces the lattice data of the chiral condensate.
Then we use the metric to holographically calculate the Wilson loop, the quark antiquark potential,
which is our prediction of the model.

In any holographic QCD, precise comparison with experiments is not expected, as
the models should be able to capture only generic property of QCD: the AdS/CFT correspondence
is valid at the strong coupling limit, and any holographic QCD model 
is not expected to be completely equivalent to QCD. So we concentrate on
how the DL method produces any novel feature of the bulk metric. 
In fact, we find that the DL modeling works beyond our imagination. The generated metric, surprisingly, has
co-existing two features: the confinement and the Debye screening.
Normally in AdS/CFT, due to the strong coupling limit and the large $N_c$ limit, the bulk spacetime is governed 
by classical Einstein gravity and thus the thermal phase transition is
described by the Hawking-Page transition \cite{Hawking:1982dh}, which is the first order.
So there is no mixture of the confinement and the Debye screening.
On the other hand, 
in our case, we simply use a lattice QCD data (which is of course not at the strong coupling limit), 
and the machine-generated metric
would capture both of the phases at the same time. Resultantly, our holographically calculated Wilson loops
share the important properties with the lattice QCD Wilson loops: the co-existence of
the  linear potential part and the Debye screening part.
Such a bulk metric has not been employed in holographic QCD modeling, and
finding this new feature is a virtue of the deep learning method.

It is often discussed in literature whether there exits any relation between the chiral symmetry breaking and the
quark confinement 
(see Refs.~\cite{Suganuma:1993ps,Miyamura:1995xn,Woloshyn:1994rv,Fukushima:2002ew,Hatta:2003ga,Gattringer:2006ci,Bilgici:2008qy,Synatschke:2008yt,Lang:2011vw,Gongyo:2012vx,Glozman:2012fj,Doi:2014zea,Suganuma:2017syi,Suganuma:2016lnt} for various study). 
Our model, with the vanishing chiral condensate in the chiral limit at a finite temperature,
has a confining part in the quark antiquark potential, thus showing the discrepancy between
the nonvanishing chiral condensate and the quark confinement.\footnote{However, 
at the same time, we can also argue that a peculiar shape of the emergent metric
could be a common origin of both of them. See Sec.~\ref{sec:5-2}.}

Another aspect we would like to emphasize about our approach is that 
the deep learning method can reconstruct the bulk starting from the data of the boundary
quantum field theory. Besides other methods of reconstructing bulk geometries, such 
as the entanglement entropy reconstruction \cite{Balasubramanian:2013lsa,Myers:2014jia},
our method utilizes physical observables of the boundary theory directly. 
Furthermore, in our analysis it is important to identify the neural network itself as a bulk geometry.\footnote{The concept has been pursued also in Ref.~\cite{You:2017guh}. For similar perspectives, see essays
\cite{Gan:2017nyt,Lee:2017skk}. Originally, identifying neural networks as discretized physical systems
has been argued in the context of spin networks \cite{Hopf, spin,spin2} (an example is the Hopfield model \cite{Hopf}), in which
the weights of the neural networks are identified as spin interactions. Here instead, our weights
are identified as metric, and thus the neural network itself may have an interpretation of a spacetime.}
In view of the quantum-information theoretic understanding of the AdS/CFT correspondence,
such as the one through MERA \cite{Swingle:2009bg}, we are providing a way to
obtain a discrete network as a gravity dual, which may shed some light on the mystery of the
origin of the AdS/CFT correspondence.

The organization of this paper is as follows. After we review our holographic modeling based on
deep learning provided in Ref.~\cite{Hashimoto:2018ftp} in Sec.~\ref{sec:2}, we apply the
AdS/DL correspondence to QCD in Sec.~\ref{sec:3}. We prepare lattice QCD data there and
describe how to convert it to the deep learning training data. Sec.~\ref{sec:4} shows the training result
of the emergent metric of the bulk geometry, and describes its physical features which are automatically generated
by the neural network. In Sec.~\ref{sec:5}, we calculate Wilson loops by using the emergent metric
obtained by the deep learning, and find that they capture nicely the features of Wilson loops
obtained in lattice QCD. Then we discuss the relation and the origin of the chiral symmetry breaking and
the quark confinement. Sec.~\ref{sec:6} is for a summary and discussions. Our Appendix \ref{app:1}
describes some details about the numerical codes of the deep learning. Appendix \ref{app:2} is
to fix the normalization of two-point functions in AdS/CFT.

\begin{figure}
\includegraphics[width=6.5cm]{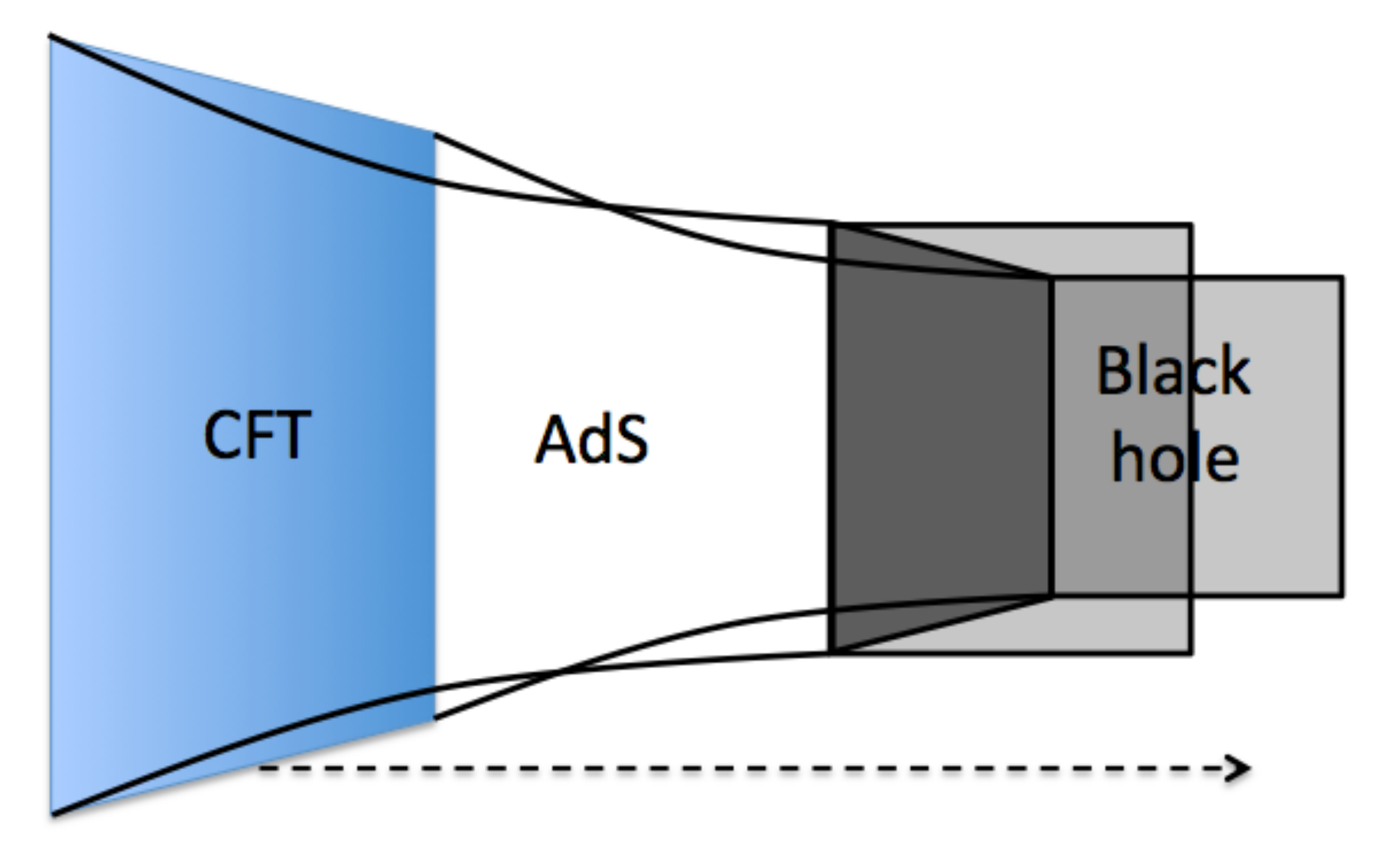}
\includegraphics[width=8.5cm]{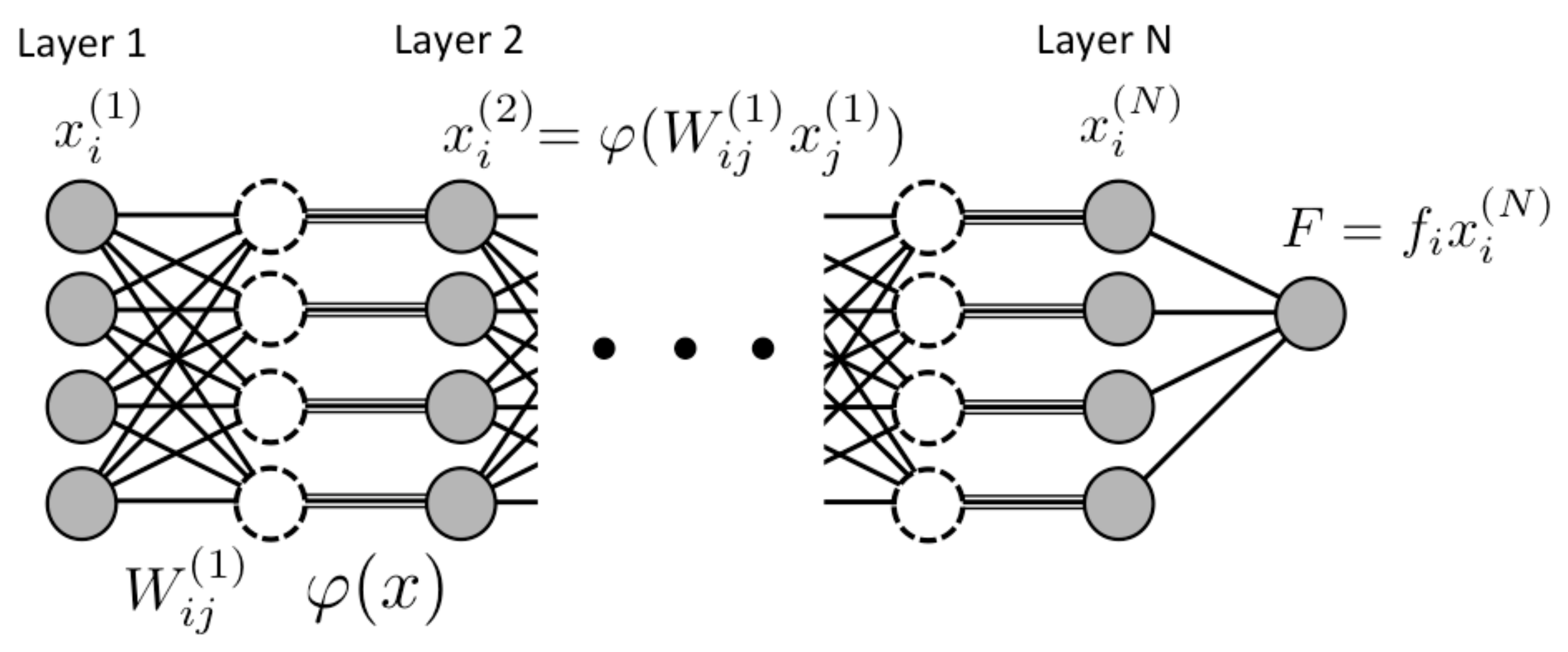}
\vspace*{-2mm}
\caption{The AdS/CFT and the DL \cite{Hashimoto:2018ftp}. 
Top: a typical view of the AdS/CFT correspondence. The CFT at 
a finite temperature lives at a boundary of asymptotically AdS spacetime
with a black hole horizon at the other end.
Bottom: a typical neural network of a deep learning.
}
\label{fig:typ}
\end{figure}

\section{Review: AdS/DL correspondence.}
\label{sec:2}

Here we review our map used in Ref.~\cite{Hashimoto:2018ftp} to relate 
a neural network to the scalar field equation in asymptotically AdS spacetime. 
The idea is to regard the depth direction of a deep neural network as the AdS radial direction
(see Fig.\ref{fig:typ}),
to have a neural network representation of the equation of motion of the 
scalar field in the curved spacetime.
Then the input data at the initial layer is a one-point function $\langle {\cal O}\rangle_J$
of the boundary quantum field theory (QFT). It is 
a function of the source $J$ for the operator ${\cal O}$. The output data 
is the black hole horizon condition for the bulk scalar field.
A supervised learning provides an emergent bulk metric which is consistent with
$\langle {\cal O}\rangle_J$.

First we briefly describe a standard deep neural network. 
It is a map from the data at the input layer to the data at the output data.
Between the input and the output layers, we prepare piles of layers,
and between the adjacent layers, the data is transmitted through 
a linear transformation $x_i \to W_{ij} x_j$ and
a nonlinear transformation $x_i \to \varphi(x_i)$.
Normally the latter, called an activation function, is a fixed function, while the former, $W$, 
called weights, are tunable parameters to be trained in the learning process.
A successive transformation among layers results in a relation between the input data $x^{(1)}$
and the output data $y$,
\begin{align}
y(x^{(1)}) =  f_i \varphi(W_{ij}^{(N-1)}\varphi(W_{jk}^{(N-2)} \cdots \varphi(W_{lm}^{(1)}x_m^{(1)}))).
\label{nns}
\end{align}
The linear transformation $f_i$ is for the final layer to wrap up the neural network to give an output number.
A supervised training is to tune the weights of the network $(f_i, W_{ij}^{(n)})$ for $n=1,2,\cdots, N-1$
so that it makes the loss function decrease, 
\begin{align}
E \equiv \sum_{\rm data}\biggm| y(\bar{x}^{(1)})-\bar{y}\biggm| + E_{\rm reg}(W).
\label{loss}
\end{align}
Once the training data $\{(\bar{x}^{(1)},\bar{y})\}$ is given, then the loss function $E$ can be calculated,
and the weights are trained so that the loss function decreases.
$E_{\rm reg}$ is a regularization term, which we later use for extracting a particular kind of weight distributions.

%
\begin{figure}
\includegraphics[width=8cm]{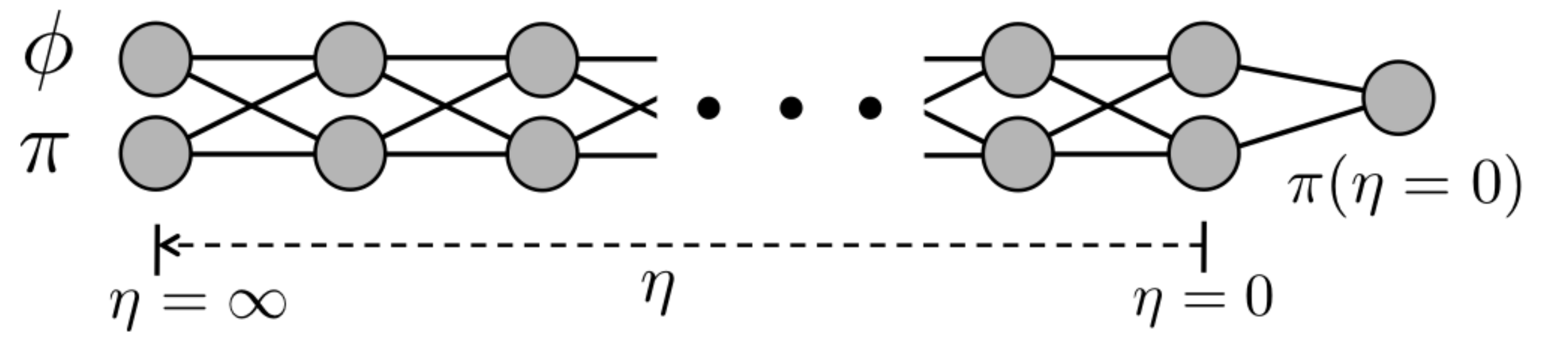}
\caption{
Our deep neural network for the emergence of the bulk metric \cite{Hashimoto:2018ftp}.
}
\label{fig:ournn}
\end{figure}
%

Next, we present a neural network representation of a scalar field equation
in an asymptotically $(d+1)$-dimensional AdS spacetime. 
We suppose that the metric takes the following form in a gauge $g_{\eta\eta}=1$: 
\begin{align}
ds^2=-f(\eta) dt^2 + d\eta^2 + g(\eta)(dx_1^2+\cdots +dx_{d-1}^2), 
\label{genericm}
\end{align}
where $\eta$ is the coordinate for the AdS radial direction $(0\leq \eta < \infty)$. 
Then, the action of a scalar field $\phi$, which is homogeneous along $t$ and $x_i$, is given by 
\begin{align}
S\!=\!VT \int\!d\eta \sqrt{f g^{d-1}}\left[
-\frac12 (\partial_\eta \phi)^2-\frac12 m^2 \phi^2-V(\phi)
\right].
\label{scalar}
\end{align}
Asymptotically $(\eta \approx \infty)$, the metric needs to be AdS, 
\begin{align}
f \approx g \approx \exp[2\eta/L+k_0] \, 
\end{align}
where  $L$ is the AdS radius and $k_0$ is a constant.
On the other hand, we assume the existence of a planer black hole horizon 
at the other side of the AdS radial direction $(\eta \approx 0)$,
\begin{align}
f \approx (2\pi T_{\rm BH})^2\eta^2 \, , 
\quad
g \approx \mbox{const.} \, ,
\label{temp}
\end{align}
where $T_{\rm BH}$ is the Hawking temperature of the black hole, corresponding to the
temperature of the boundary QFT.

The classical equation of motion for $\phi(\eta)$ is
\begin{align}
\partial_\eta \pi + h(\eta)\pi -m^2\phi- \frac{\delta V[\phi]}{\delta \phi} = 0, \hspace{5mm} \pi \equiv \partial_\eta \phi \, ,
\label{sceq}
\end{align}
where the metric function is
\begin{align}
h(\eta)\equiv \partial_\eta \log\sqrt{f(\eta)g(\eta)^{d-1}}.
\end{align}
We discretize the $\eta$ axis, then
\begin{align}
&\phi(\eta\!+\!\Delta \eta) = \phi(\eta) + \Delta\eta \, \pi(\eta)\, , 
\label{prop}
\\
&\pi(\eta\!+\!\Delta \eta) = \pi(\eta) 
- \Delta\eta 
\left(h(\eta)\pi(\eta) -m^2\phi(\eta)- \frac{\delta V(\phi)}{\delta \phi(\eta)}\right).
\label{prop2}
\end{align}
This Hamilton-like form \eqref{prop}\eqref{prop2} is in fact a neural network representation
of the bulk AdS scalar field equation. We interpret the weights as a metric,
\begin{align}
W^{(n)} = \left(
\begin{array}{cc}
1 & \Delta \eta \\
\Delta\eta \, m^2 & 1-\Delta \eta \,h(\eta^{(n)}) 
\end{array}
\right)\, ,
\label{W}
\end{align}
and the activation function as the interaction term\footnote{
Interestingly, there exists a very similar structure in a recent deep learning architecture \cite{nice}.
},
\begin{align}
\left\{
\begin{array}{l}
\varphi(x_1) = x_1,
\\ \varphi(x_2) = x_2 + \Delta \eta \, \displaystyle\frac{\delta V(x_1)}{\delta x_1} \, .
\end{array}
\right.
\label{act}
\end{align}
These bring \eqref{prop}\eqref{prop2} to the form \eqref{nns}, once we interpret the
vectors $x_i$ as $(\phi,\pi)$ and the layer index as the discritezed $\eta$. See Fig.~\ref{fig:ournn}.

The input data is the pair $(\phi(\eta=\infty),\pi(\eta=\infty))$. The values are obtained once the one-point function
of a QFT operator ${\cal O}$ is given under the source $J$ in the boundary QFT. 
In fact, $\langle {\cal O}\rangle_J$ and $J$ correspond to coefficients of the 
normalizable and non-normalizable modes of $\phi$ in the asymptotically AdS spacetime
\cite{Klebanov:1999tb}, and later we will provide an explicit formula for relating these for the case of QCD chiral
condensate.
For numerical simulations, we introduce a cutoff $\eta= \eta_{\rm ini}$ and regard that point as the location of
the asymptotic AdS boundary.
 
On the other hand, at $\eta=0$, there is a black hole horizon and we put another regularized cutoff $\eta=\eta_{\rm fin}$ close to $\eta=0$. There, if the bulk scalar field consistently satisfies the
in-going boundary condition, then we need to require (see for example \cite{Horowitz:2010gk})
\begin{align}
0=F\equiv \left[\frac{2}{\eta} \pi -m^2 \phi - \frac{\delta V(\phi)}{\delta \phi}\right]_{\eta=\eta_{\rm fin}}
\label{Ff}
\end{align}
In the limit $\eta_{\rm fin}\to 0$, the condition \eqref{Ff} is equivalent to $\pi(\eta=0)=0$,
so this is the output data of the neural network.

The training of the neural network is given by these positive data, $(\phi(\eta=\infty),\pi(\eta=\infty))$ as
the input and $\pi(\eta=0)=0$ as the output. For the training we also need negative data which can be
generated easily by looking at the data points away from the positive data. 
We may assign $\pi(\eta=0)=1$ for the negative data for convenience. 

\begin{figure}
\begin{center}
\includegraphics[width=8.5cm]{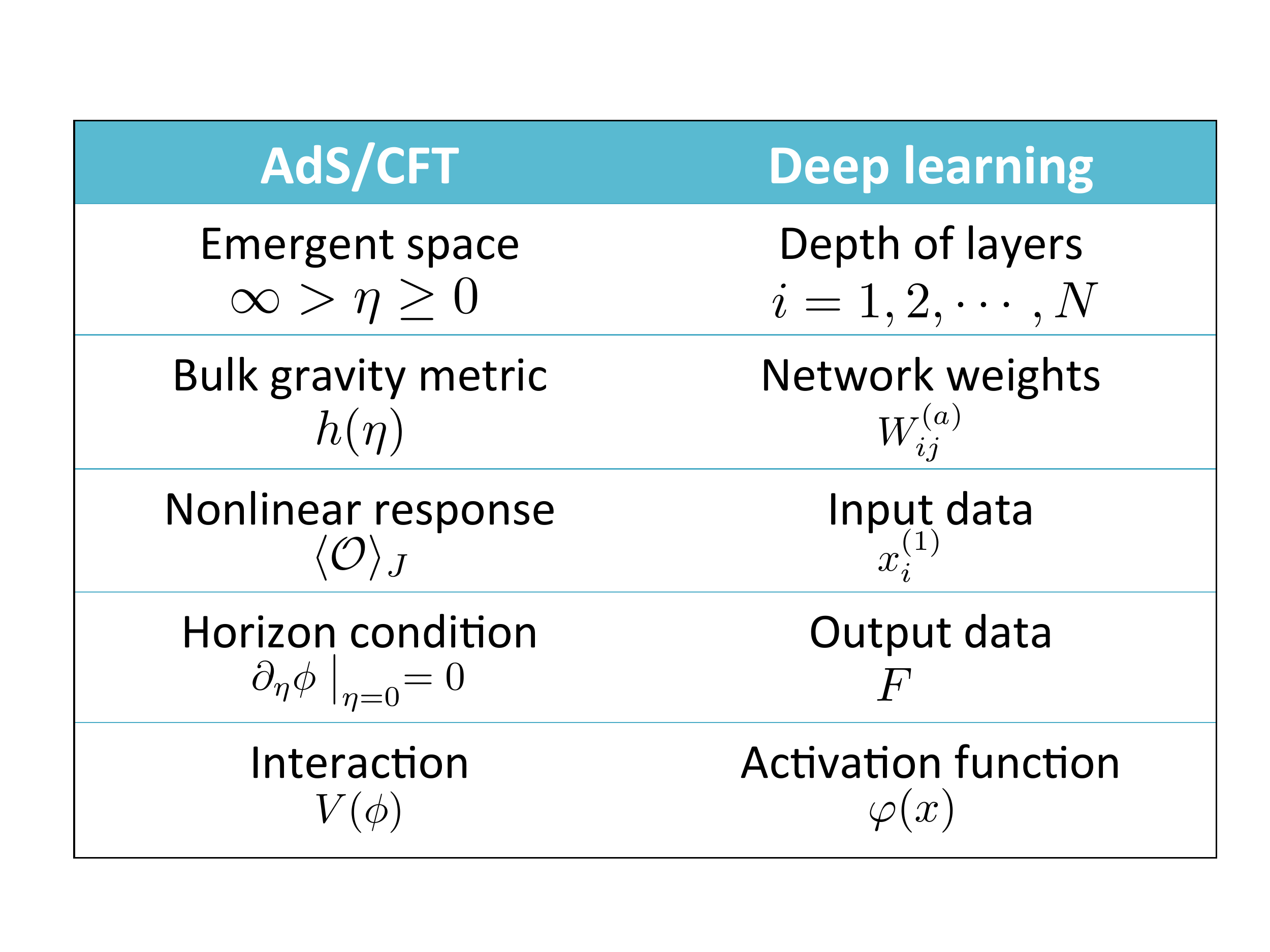}
\end{center}
\vspace{-5mm}
\caption{
Our dictionary between the AdS/CFT correspondence and the deep learning.}
\label{fig:ourdic}
\end{figure}

In summary, the dictionary between the AdS/CFT correspondence and the deep learning method, specifically for the bulk scalar field equation in the background of black hole geometry, is given in Fig.~\ref{fig:ourdic}.
The most important aspect of the dictionary is the fact that 
the weights in the neural network correspond to the bulk geometry.
See appendix A for details of the numerical training of the machine learning.

%
%

\section{Lattice QCD data as input.}
\label{sec:3}

As explained in the introduction, one of the virtue of our holographic modeling with use of the
deep neural network is to construct a gravitational metric from a QCD data.
We use the input data as the one-point function of a QCD operator under a source $J$,
and the simplest and most popular example of such in QCD is the chiral condensate 
${\cal O}\equiv \bar{q}q$.
The source for the chiral condensate is the quark mass $m_q$, since in the QCD Lagrangian 
the quark mass term is written as $m_q {\cal O}$.

There are various lattice QCD data for the chiral condensate, among which we chose a data 
of RBC-Bielefeld collaboration \cite{Unger} because of the following two reasons: First, the data 
is taken near the critical temperature of QCD, and second, the quark mass dependence 
and its temperature dependence is clearly interpreted from the data.
Since our output data is the black hole boundary condition, the lattice data needs to be
that above the thermal phase transition temperature. Normally, AdS/CFT correspondence is
studied at a strong coupling limit, and at the limit the thermal phase transition is the first order.
On the other hand, the thermal phase transition of QCD is crossover. So it is interesting
to ask our neural network what is the emergent metric out of the lattice data near the phase
transition temperature. We will see a surprising consequence in later sections.

\begin{table} 
\begin{tabular}{|c|c|}
$m_q$  & $\langle \bar{\psi}\psi\rangle$ \\
\hline
0.0008125 & 0.0111(2)\\
0.0016250 &0.0202(4) \\
0.0032500 & 0.0375(5)\\
0.0065000 & 0.0666(8)\\
0.0130000 & 0.1186(5) \\
0.0260000 & 0.1807(4)
\end{tabular}
\hspace{5mm}
\begin{tabular}{|c|c|}
$m_q$[GeV]  & $\langle \bar{\psi}\psi\rangle$ [(GeV)$^3$]\\
\hline
0.00067 &0.0063
\\
0.0013 &0.012
\\
0.0027 &0.021
\\
0.0054 &0.038
\\
0.011 &0.068
\\
0.022 &0.10
\end{tabular}
\caption{Left: Lattice data of the chiral condensate as a function of quark mass \cite{Unger},
in the unit of lattice spacing $a$. Right: its translation to the physical units, using $1/a =0.829(19)$[GeV].}
\label{latticedata}
\end{table}

The lattice QCD data \cite{Unger} for the chiral condensate as a function of the quark mass is shown in Table \ref{latticedata}.\footnote{We would like to thank W.~Unger for providing us with the lattice data of the RBC-Bielefeld collaboration.}
The unit is the lattice spacing $a$, and the data is for the QCD coupling constant $\beta_\text{lat} = 6/g_{\rm lat}^2 = 3.3300$,
where $g_\text{lat}$ is the gauge coupling constant at the cutoff scale $a^{-1}$, namely $g_\text{lat} = g(\mu= a^{-1})$.

Here we briefly review the determination of the lattice spacing $a$ in the physical scale \cite{Sommer:1993ce}.
First of all, the cutoff $a^{-1}$ and the gauge coupling $\beta_\text{lat}=6/g^2(\mu= a^{-1})$ are related by the QCD beta function,
\begin{align}
\mu \frac{d g(\mu)}{d \mu} = - b_0 g(\mu)^3 + \cdots
\end{align}
where $\mu$ is a scale and $b_0$ is the first coefficient of the QCD beta function.
By solving this differential equation, we can determine the relation between $\beta_\text{lat}$ and the cutoff $a^{-1}$.
However, we cannot fix the absolute energy scale, as in the perturbative renormalization procedure, therefore
we need physical observables to fix it. 
R.~Sommer introduced the heavy quark potential to fix the scale \cite{Sommer:1993ce},
\begin{align}
r^2\frac{\partial }{\partial r} V(r)\bigg|_{r=r_0} = 1.65, \label{eq:sommer_scale}
\end{align}
where $V(r)$ is the potential energy between two heavy quarks, 
which is calculated using Wilson loops on the lattice. The left hand side is a dimensionless combination.
A constant in the right hand side is chosen such that $r_0$ is equal to $0.469(7) \approx 0.5$ [fm]. 
By measuring the left hand side of \eqref{eq:sommer_scale} on the lattice,
we can determine a dimensionless number $r_0/a$, and using $r_0 \approx 0.5$ [fm],
we can determine the absolute scale $a$ in the physical unit.

In Ref.~\cite{scale}, they determined $r_0/a=1.995(11)$ for $\beta_\text{lat}=3.335$ and $r_0/a=1.823(16)$ for $\beta_\text{lat}=3.290$
for dynamical two-flavor staggered QCD.
Since the above data for the chiral condensate is for  $\beta_\text{lat}=3.3300$, we interpolate the two values of
$r_0/a$ and obtain $r_0/a=1.976(16)$ for $\beta_\text{lat}=3.3300$. Using $r_0=0.469(7)$ [fm], we obtain
$a=0.238(6)$[fm],
which means $1/a = 829(19)$ [MeV].
This value is used to translate the lattice simulation values in the left panel of
Table \ref{latticedata} to their physical values,
as given in the right panel of Table \ref{latticedata}.
It is also used to determine the temperature $T=1/(4a)= 207(5)$ [MeV], where the factor $4$ is
for  the temporal size of the used lattice in the lattice unit $a$. 
The value of the temperature shows that the input lattice data is around the critical temperature
of the QCD thermal phase transition.

\begin{figure}
\includegraphics[width=8.5cm]{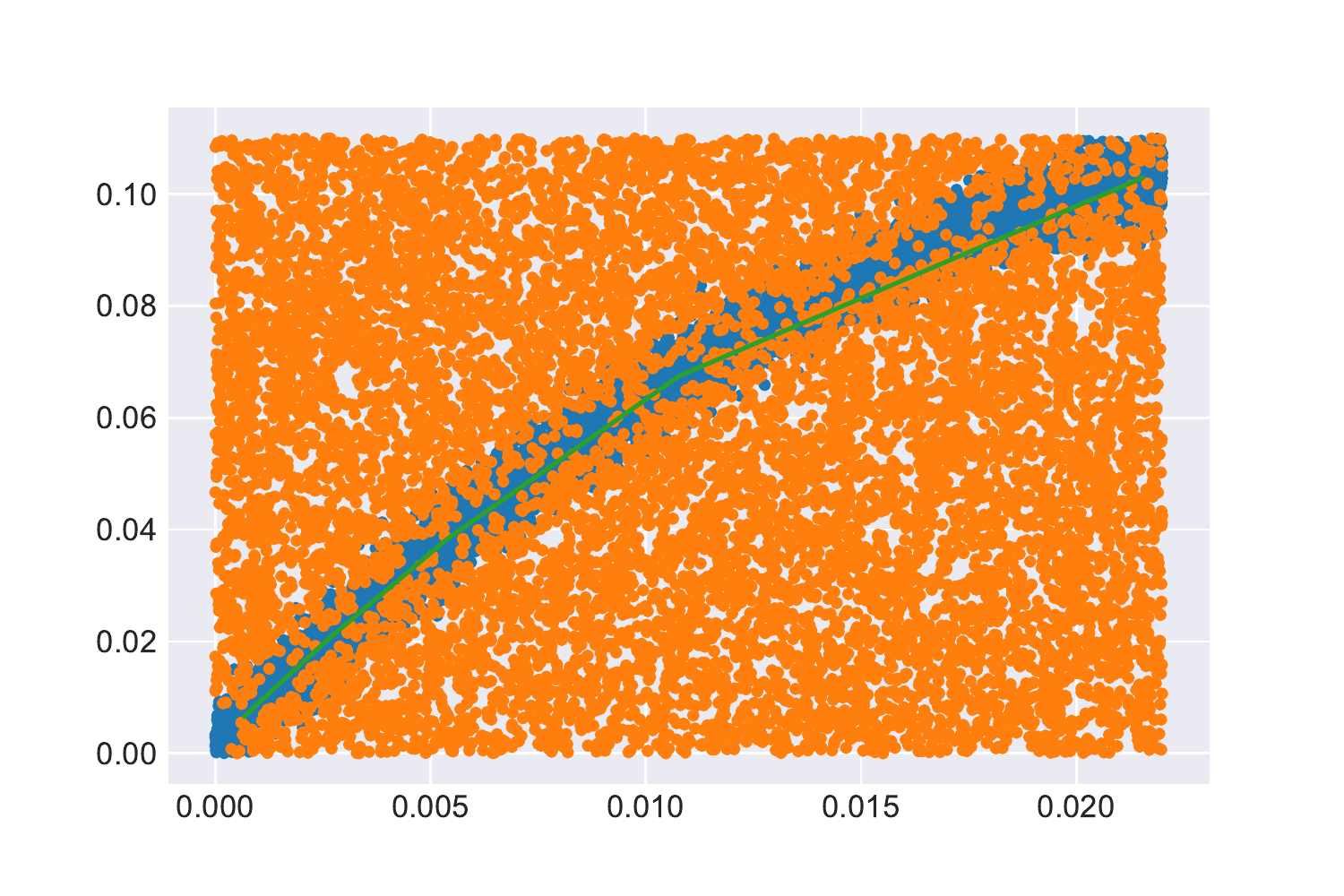}
\caption{Data used for training the neural network. The horizontal axis
is the quark mass [GeV], and the vertical axis is the chiral condensate
[GeV${}^3$].
Blue dots are positive data, while orange dots are negative data.
}
\label{fig:data}
\end{figure}

Using the lattice QCD data of $\langle{\cal O}\rangle_J=\langle \bar{q}q\rangle_{m_q}$ in physical units, described in the right panel of
Table \ref{latticedata}, we may generate a data set to train the neural network.
Since the error expected in converting the lattice data to physical units is at most 8 $\%$ 
(which mainly comes from the error of the value of $1/a^3$), we take it into account for
generating the training data set. The generated data set which we use is shown in Fig.~\ref{fig:data}.

We generate the positive and negative data as follows. First, using the lattice
data of the right panel of Table \ref{latticedata}, we fit it with a polynomial function up to the third power
in $m_q$. Then we randomly pick a point in the region $0<m_q<0.022$ and $0<\langle \bar{q}q\rangle < 0.11$
and judge whether it belongs to the positive data set or the negative data set by checking whether 
the vertical distance from the fitted curve is less than or more than 0.004. We collect 10000 positive data points
and 10000 negative data points. They are plotted in Fig.~\ref{fig:data}.

Next, we need to map the generated data to the first layer of the neural network by using 
the map given in Ref.~\cite{Klebanov:1999tb}.
We consider a $\phi^4$ theory in the bulk, and take
\begin{align}
V[\phi]= \frac{\lambda}{4} \phi^4 \, .
\end{align}
The coupling constant $\lambda$ is to be trained, and it should be positive $(\lambda > 0)$.
Furthermore, the chiral condensate $\langle \bar{q}q\rangle$ has mass dimension 3 at the UV Gaussian fixed point,
so we have to take $m^2 = -3/L^2$, according to the  well-known relation between the 
bulk scalar mass and the conformal dimension $\Delta_{\cal O}$ of the operator ${\cal O}$ 
in $d$ spacetime dimensions,
\begin{align}
\Delta_{\cal O} \equiv (d/2)+\sqrt{d^2/4 + m^2L^2} \, .
\label{DeltaO}
\end{align}
At the asymptotic boundary of the geometry, the spacetime is the AdS with the AdS radius $L$.
This means $h(\eta) \approx 4/L$. 
Then the bulk equation of motion is written near the asymptotic boundary as
\begin{align}
\partial_\eta^2 \phi + \frac{4}{L}\partial_\eta \phi -\frac{3}{L^2}\phi - \lambda \phi^3 = 0 \, .
\end{align}
The solution of the scalar field equation near the asymptotic AdS spacetime is
\begin{align}
L^{3/2}\, \phi  \approx
\alpha e^{-\eta/L} + \beta e^{-3\eta/L} -\frac{\lambda\alpha^3}{2L^2} \eta \, e^{-3\eta/L}.
\end{align}
The first two terms are non-normalizable and normalizable modes, corresponding to
the quark mass and the chiral condensate, according to the AdS/CFT dictionary.
The third term is present and necessary, 
when the conformal dimension is an integer. 
In numerical simulations we deal with dimension-less quantities,
and everything is measured in units of the AdS radius $L$.
The normalization of dimensionless $\alpha$ and $\beta$ is determined (see Appendix~\ref{app:2}),
\begin{align}
\alpha = \frac{\sqrt{N_c}}{2\pi} m_q L \, , \quad
\beta = \frac{\pi}{\sqrt{N_c}} \langle \bar{q}q\rangle L^3 \, .
\label{alb}
\end{align}
We take $N_c=3$, and the value of the AdS radius, $L$, 
needs to be trained in the machine learning procedures.

In summary, the input data for the neural network is given by
\begin{align}
&
\phi(\eta_{\rm ini}) = 
\alpha e^{-\eta_{\rm ini}} + \beta e^{-3\eta_{\rm ini}} -\frac{\lambda\alpha^3}{2} \eta_{\rm ini} \, e^{-3\eta_{\rm ini}}.
\label{phiini}
\\
&\pi(\eta_{\rm ini}) = 
-\alpha e^{-\eta_{\rm ini}} -\left(3\beta+\frac{\lambda\alpha^3}{2}\right) e^{-3\eta_{\rm ini}} 
\nonumber 
\\
&
\hspace{40mm}
+\frac{3\lambda\alpha^3}{2}  \eta_{\rm ini}e^{-3\eta_{\rm ini}},
\label{piini}
\end{align} 
with the coefficients \eqref{alb} with $N_c=3$, with the positive and negative data given in Fig.~\ref{fig:data}.
The variables to be trained are: $\lambda$, $L$ and $h(\eta)$.

\section{Deep learning and emergent metric.}
\label{sec:4}

\subsection{Preparation of the neural network.}

First, let us discretize the $\eta$ direction. We prepare $N=15$ layers,
and the UV and IR cutoffs are introduced as $\eta_{\rm ini}=1$ and 
$\eta_{\rm fin}=0.1$, in units of $L$. So we discretize the region $\eta_{\rm fin} \leq
\eta \leq \eta_{\rm ini}$ to 15 points which are apart equidistantly.
Note that the $\eta$ cutoff values
are not relevant except for the difference $\eta_{\rm ini}-\eta_{\rm fin}$,
because the bulk equations of motion \eqref{prop} \eqref{prop2} are invariant under translation
along $\eta$. We will see later that this arbitrariness is used to fit the emergent metric
with the location of the black hole horizon.

Since we are interested in a smooth continuum limit of $h(\eta)$ which asymptotes to the AdS spacetime,
we employ the following two regularization 
terms \footnote{Note that
we did not introduce a regularization corresponding to the horizon behavior. Such a 
regularization was used in our previous 
paper \cite{Hashimoto:2018ftp}.
The divergence of the metric function $h(\eta)\sim 1/\eta$ is automatically obtained through the training in 
the present case.}, $E_{\rm reg} =E_{\rm reg}^{\rm (smooth)}+E_{\rm reg}^{\rm (bdry)}$. The first one is 
\begin{align}
E_{\rm reg}^{\rm (smooth)} 
\equiv c_{\rm reg}\sum_{n=1}^{N-1}(\eta^{(n)})^4 \left(h(\eta^{(n+1)})-h(\eta^{(n)})\right)^2
\end{align}
with $c_{\rm reg}=0.01$. With this regularization, the weights $h(\eta)$ which are smooth in $\eta$ are favored.
The factor $\eta^4$ is introduced such that the regularization also allows a functional form $h(\eta)\propto 1/\eta$ near
$\eta \approx 0$. In the continuum limit $N\to\infty$, this regularization is roughly equivalent to $c_{\rm reg} \int d\eta\, (h'(\eta)\eta^2)^2$.
The value of $c_{\rm reg}$ is chosen such that the training rejects zigzag-shaped metric functions 
and at the same time the training proceeds smoothly. The second one is 
\begin{align}
E_{\rm reg}^{\rm (bdry)} \equiv c_{\rm reg}\left(d-h(\eta^{(1)})\right)^2
\end{align}
with the QFT spacetime dimension $d=4$. Since for the pure AdS spacetime we have $h(\eta)=d$, 
at the asymptotic region $\eta=\eta^{(1)}=\eta_{\rm ini}$ our metric needs to be smoothly connected to
$h=d$. This regularization is to choose weights which are consistent with the asymptotically AdS spacetime.
Again, for simplicity, we choose $c_{\rm reg}=0.01$.

In our numerical training, we use PyTorch for a Python deep learning library to implement our network.
The trained variables are 15 values of $h(\eta^{(n)})$, the coupling constant $\lambda$ and the AdS radius $L$.
The initial metric function is randomly chosen (the standard deviation of magnitude 3 with the mean 
value $h=4$), and the initial values of $\lambda$ and $L$ are chosen as $\lambda = 0.2$ and 
$L=0.8$[GeV${}^{-1}$]. 
In the learning process, we choose the batch size equal to 10, and the stop training at 1500 epochs.

\subsection{Obtained metric function.}

\begin{figure}
\begin{center}
\includegraphics[width=8.5cm]{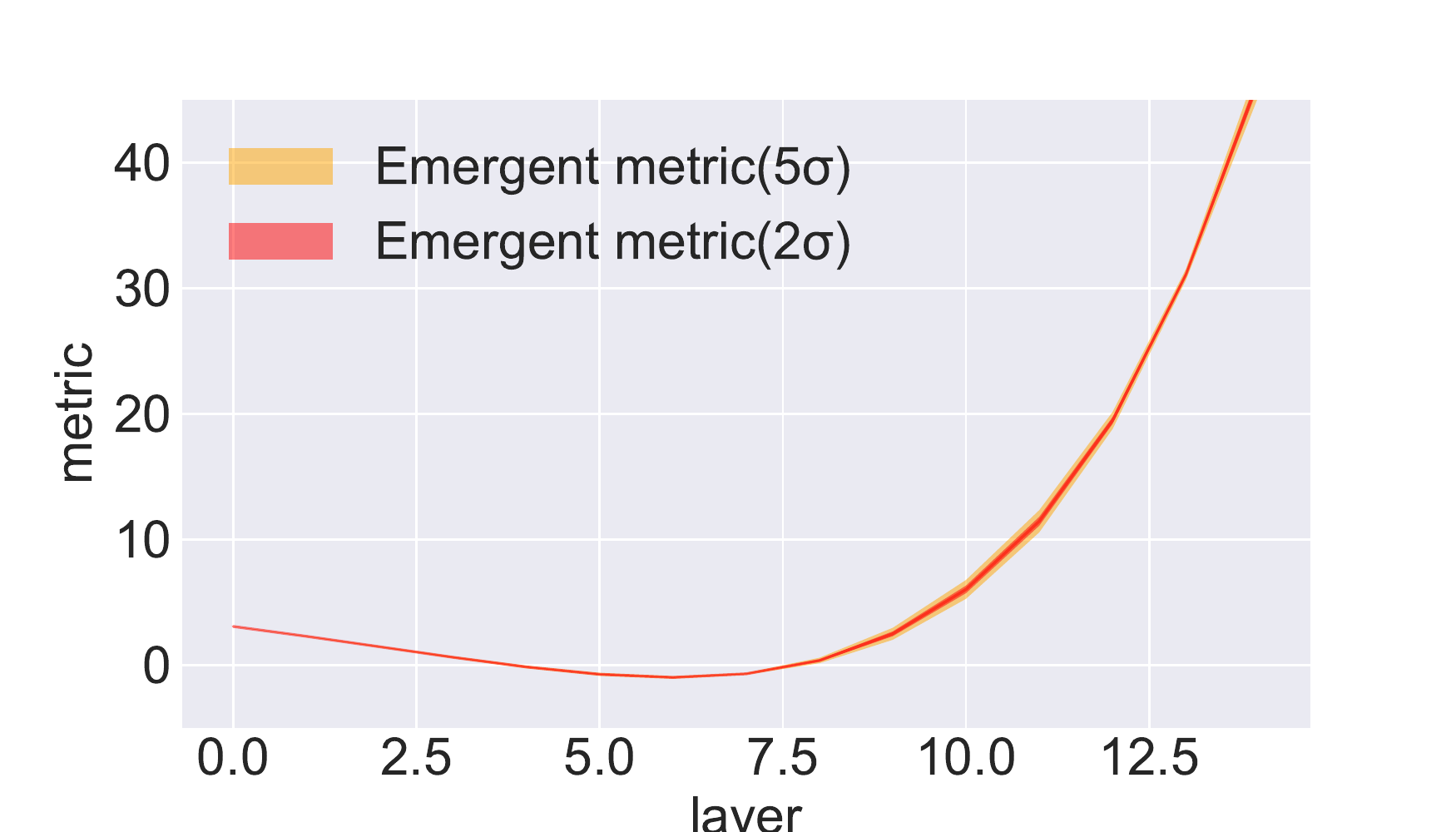}
\end{center}
\caption{Plot of the trained metric function $h(\eta)$ given in Table \ref{metrictable}. 
The horizontal axis is the layer index $n-1=0,\cdots,14$.}
\label{fig:metric}
\end{figure}

We collect 8 statistical data under the criterion that the total loss after the 1500 epochs is less than 0.08.\footnote{
The loss value 0.08 is chosen since we numerically have not observed any training which goes beyond that: we terminate the training when the loss does not decrease.}
The statistical result of the metric function $h(\eta^{(n)})$ is shown in Table \ref{metrictable},
and its plot is given in Fig.~\ref{fig:metric}. We also obtained the trained values of the coupling constant and the AdS radius,
\begin{align}
& \lambda = 0.01243 \pm 0.00060 \, , 
\\
& L  = 3.460 \pm 0.021 \mbox{[GeV${}^{-1}$]} \, . 
\end{align}
Using 5.0677 [GeV${}^{-1}$] $=$ 1 [fm], our AdS radius is determined as
$L=0.6828 \pm 0.0041$ [fm].

\begin{table} 
\begin{tabular}{|c|c|}
$n-1$  & $h$ \\
\hline
0 & $3.0818 \pm 0.0081$\\
1 & $2.296 \pm 0.016$\\
2 & $1.464 \pm 0.025$\\
3 & $0.627 \pm 0.035$\\
4 & $-0.141 \pm 0.045$\\
5 & $-0.727 \pm 0.049$\\
6 & $-0.974 \pm 0.043$\\
7 & $-0.687 \pm 0.032$\\
8 & $0.374 \pm 0.087$\\
9 & $2.50 \pm 0.19$\\
10 & $6.03 \pm 0.30$\\
11 & $11.46 \pm 0.35$\\
12 & $19.47 \pm 0.27$\\
13 & $31.07 \pm 0.17$\\
14 & $46.70 \pm 0.52$\\
\end{tabular}
\caption{Values of the trained metric function $h(\eta^{(n)})$ for the layer index $n-1=0,1,\cdots,14$.
We collected 8 metrics and
the standard deviation of them is also given. }
\label{metrictable}
\end{table}

%

There are three important aspects of the obtained metric function $h(\eta)$. 
\begin{itemize}
\item
All 8 metrics converge to almost the same curve, so there is a universality of the obtained metric functions.
\item
The obtained $h(\eta^{(n)})$ diverges when $\eta$ approaches the horizon $n=15$. This is consistent with
the blackhole horizon behavior $h\approx 1/\eta$ expected from Einstein equations. Note that
we have not introduced any regularization to force the metric to prefer the divergence. The neural network
automatically captures the horizon behavior from the training data.
\item
The obtained metric function $h(\eta)$ goes negative in the middle region. This behavior is unexpected,
and it turns out that this is quite an important new feature which the machine automatically figures out,
as we will explain below.
\end{itemize}
All of these properties are important and point a successful training of the metric function in our deep learning
holographic modeling. 

\begin{figure*}
\begin{center}
\includegraphics[width=18cm]{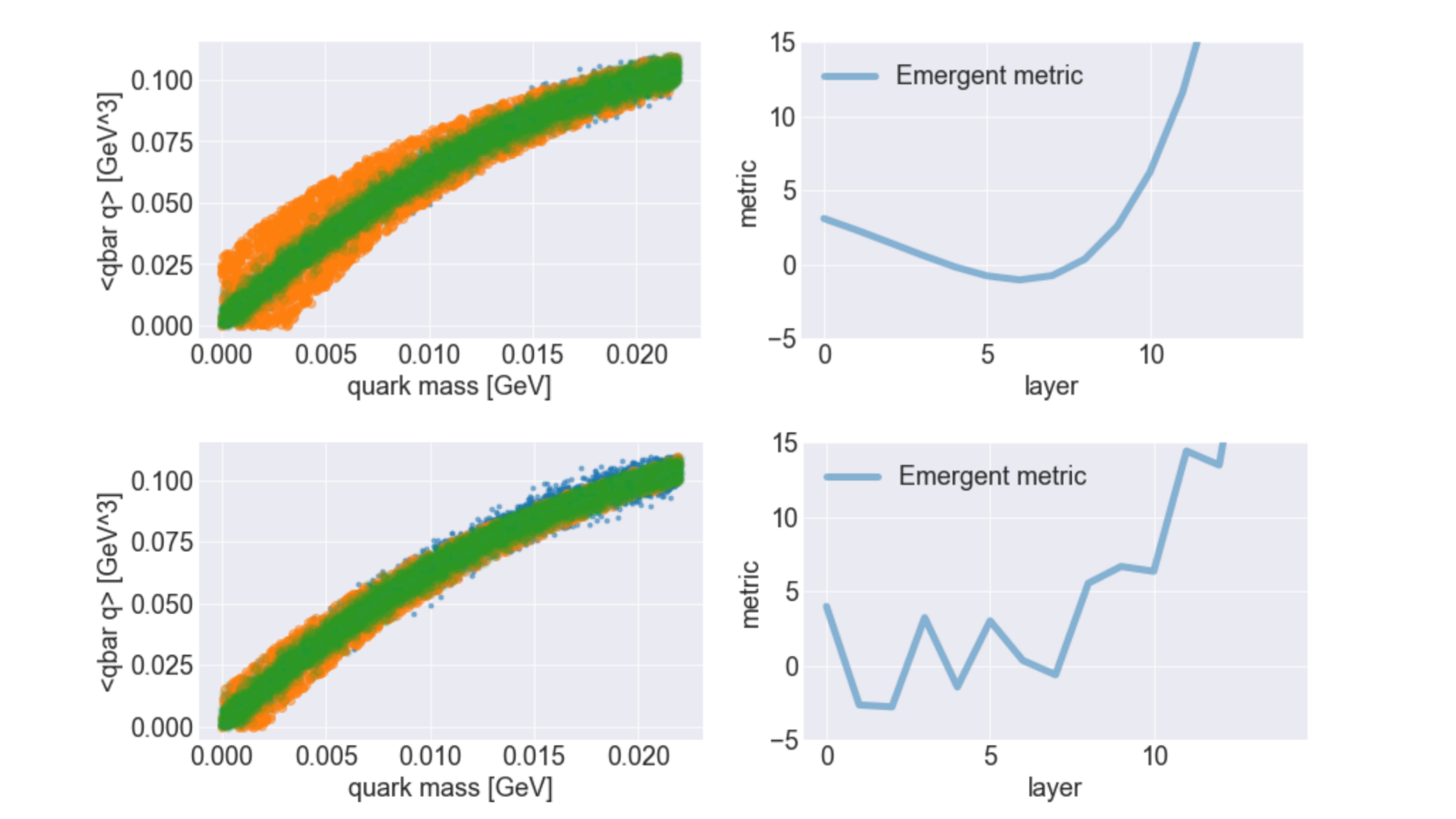}
\end{center}
\vspace{-5mm}
\caption{Data reproduction and the trained metrics. Top left: Reproduced data of one-point function. Green dots are original data, while orange + green dots are the data judged as positive by the neural network with the regularization. Top right: The emergent metric function $h(\eta)$. The loss is 0.073. Bottom: The trained metric and data without the regularization. The metric function is not smooth, while the loss goes down to 0.0079. }
\label{fig:result}
\end{figure*}

To emphasize the importance of the regularization, we show in Fig.~\ref{fig:result} the two representative 
cases of the training. The upper panels show the training with the regularization 
$E_{\rm reg} =E_{\rm reg}^{\rm (smooth)}+E_{\rm reg}^{\rm (bdry)}$, while the lower panels show that with only the 
second regularization $E_{\rm reg} =E_{\rm reg}^{\rm (bdry)}$. We find that without the
first regularization $E_{\rm reg}^{\rm (smooth)}$ the obtained metric function is not smooth.

%

However note also that the training without the first regularization $E_{\rm reg}^{\rm (smooth)}$ get smaller loss, $E=0.0074$,
compared to the case of the full regularization, $E=0.094$. The difference is apparent in the left two panels
which show the extent of the coincidence between the original lattice data (green dots) and the ones judged as
positive data by the trained metric function (orange + green dots). The upper left panel has a broader
coincidence while the lower left panel (which is without $E_{\rm reg}^{\rm (smooth)}$) has a better coincidence.
This is apparently due to the difference of the regularizations. Since we look for a gravity dual which 
has a smooth metric, we are led to conclude that we have to allow a broader coincidence of the data
to have a gravity dual.\footnote{The other way to interpret the results is that if we require a perfect data matching
then we find no good gravity model with a completely smooth metric function. This is always expected,
since QCD is not a strong coupling limit at which a gravity dual is described by Einstein gravity.
Infinite number of derivatives in gravity are expected at a finite coupling of the boundary theory,
allowing generically a zig-zag configuration of the bulk metric.
Note that we always regard $N_c=3$ as $N_c=\infty$ in holographic models.
}

\subsection{Reconstruction of full metric.}

The metric function obtained by the training of the neural network is $h(\eta)\equiv \partial_\eta \left[\log\sqrt{f(\eta)g(\eta)^3}\right]$, and in order to
calculate other physical quantities in the holographic model we need to reconstruct components of the metric, $f(\eta)$
and $g(\eta)$, from the obtained $h(\eta)$. 
There are two obstacles in getting them; first, the function $h(\eta)$ includes a derivative,
so in the integration there appears an integration constant. Second, the deep learning gives only a combination
$f(\eta) g(\eta)^3$, so one need to assume $g(\eta)$ to obtain $f(\eta)$, for example.
Fortunately the first obstacle can be fixed by requiring the temperature. 
The lattice data used for our training is at the temperature $T = 207(5)$ [MeV], and we can use it to determine
the integration constant by demanding the consistency between the temperature value and the near horizon metric,
see \eqref{temp}. 

Let us describe our procedure to find the metric components $f(\eta)$ and $g(\eta)$ consistent with 
the trained $h(\eta)$. We describe everything in units of $L$ below.
Write the $g(\eta)$ component as
\begin{align}
g(\eta) = \exp[2\eta + k(\eta)]\, .
\label{gdef}
\end{align}
For this to be consistent with the asymptotically AdS spacetime, we require
\begin{align}
k(\eta=\infty) = k_1
\end{align}
which is a constant. We also define $k_0 \equiv k(\eta=0)$.
With the obtained metric function $h(\eta)$ we define its deviation from the AdS Schwartzschild metric,
\begin{align}
H \equiv h-4\coth[4\eta] \, .
\end{align}
From the definition, $f$ and $g$ satisfies 
\begin{align}
f g^3 = \exp\left[2\int_c^\eta \left(H(\eta')+4\coth [4\eta']\right) d\eta'\right] \, .
\end{align}
Here $c$ is the integration constant. Together with \eqref{gdef}, we obtain
\begin{align}
\frac{f}{e^{2\eta}}=\exp\left[
2 \log \frac{\sinh [4\eta]}{\sinh [4c]} -8\eta -3k + 2\int_c^\eta \!\!\! H d\eta'
\right].
\end{align}
With this expression, we require two consistency conditions; first, we require that
asymptotically ($\eta \to \infty)$ the spacetime is AdS, that is, $f \approx \exp[k_1 + 2\eta]$.
Then we obtain
\begin{align}
k_1 = -\frac12 \log\left[2 \sinh [4c]\right] + \frac12 \int_c^\infty \!\!\! H(\eta') d\eta'.
\label{k1c}
\end{align}
Second, we require the temperature condition for $f$, which is the first equation of \eqref{temp}. This leads to
\begin{align}
k_0 = \frac{2}{3}\int_c^0 \!\!\! H(\eta') d\eta' - \frac23\log\left[
\frac{\pi}{2}T_{\rm BH} L \sinh[4c]
\right].
\label{k0c}
\end{align}
These two equations, \eqref{k1c} and \eqref{k0c}, are necessary conditions for $g$ to satisfy.

\begin{figure}
\begin{center}
\includegraphics[width=7.5cm]{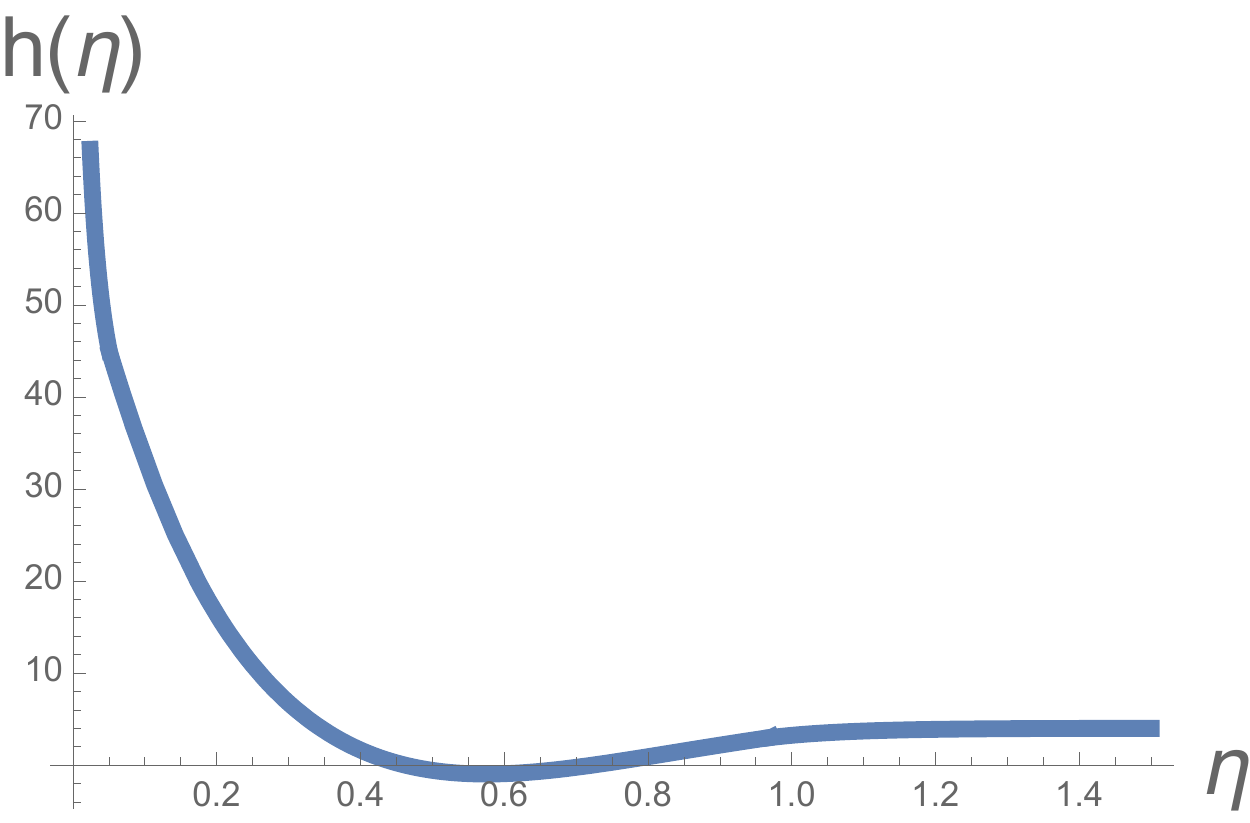}
\end{center}
\vspace{-5mm}
\caption{The fitted $h(\eta)$ which has the desired asymptotic behavior: $h(\eta)\approx 1/\eta$ near $\eta\approx 0$
and $h(\eta)\approx 4$ at $\eta\to\infty$, while reproducing the trained values of Table \ref{metrictable} in the region
$1/15-c_1<\eta<1-c_1$.}
\label{fig:h}
\end{figure}
%

The choice of $g(\eta)$, which is the choice of $k(\eta)$, cannot be fixed in our learning approach, and we choose it by hand. Our natural choices
are (i) $g(\eta) = \alpha\, e^{2\eta}$, and (ii) $g(\eta) = \alpha \cosh [2\eta]$. The latter case (ii) is the form for
AdS Schwartzschild black hole. For given $k(\eta)$, one can solve \eqref{k1c} and \eqref{k0c},
to obtain the integration constant $c$. For example, for the case (ii), from \eqref{k1c} and \eqref{k0c} we obtain
\begin{align}
\log\sinh[4c]+\int_0^c \!\! H d\eta' +3\int_0^\infty \!\! Hd\eta'=
\log\frac{2}{(\pi T_{\rm BH} L)^4}.
\end{align}
Solving this equation numerically gives the explicit value of $c$.

Now, let us fit the numerical data of Table \ref{metrictable} and obtain a smooth function $h(\eta)$.
We first fit the data in the region $\eta_{\rm fin}=1/15\leq \eta\leq \eta_{\rm ini}=1$ by a 5th order
polynomial. Then we extrapolate the function to the regions $0<\eta<\eta_{\rm fin}$ and $\eta_{\rm ini}<\eta$.
As we mentioned, the origin of the $\eta$ axis can be shifted slightly, and since we need to require that
$h(\eta)\approx 1/\eta$ near $\eta\approx 0$ which is the black hole horizon condition, 
we fit the values of the polynomial-fit $h$ near the final layer $1/15<\eta<1/15+0.05$
as $h(\eta) \approx 1/(\eta^{(n)}-c_1)+c_2$. This determines $c_1=0.0227$ and $c_2=25.257$.
It means that $\eta$ should be shifted by the small value $c_1$. 
With this shifted $\eta$, we also fit the values near the initial layer $1-c_1-0.05<\eta<1-c_1$ as 
$h(\eta)\approx 4-c_3 e^{-c_4 \eta}$.
This determines $c_3=8.248\times 10^{3}$ and $c_4=9.280$.
Using these variables, we obtain the fitted $h(\eta)$ for all the region of $0<\eta<\infty$, with the 
desired asymptotic behavior at the black hole horizon and the asymptotic AdS spacetime. See
Fig.~\ref{fig:h} for the plot of $h(\eta)$.

\begin{figure}
\begin{center}
\includegraphics[width=7.5cm]{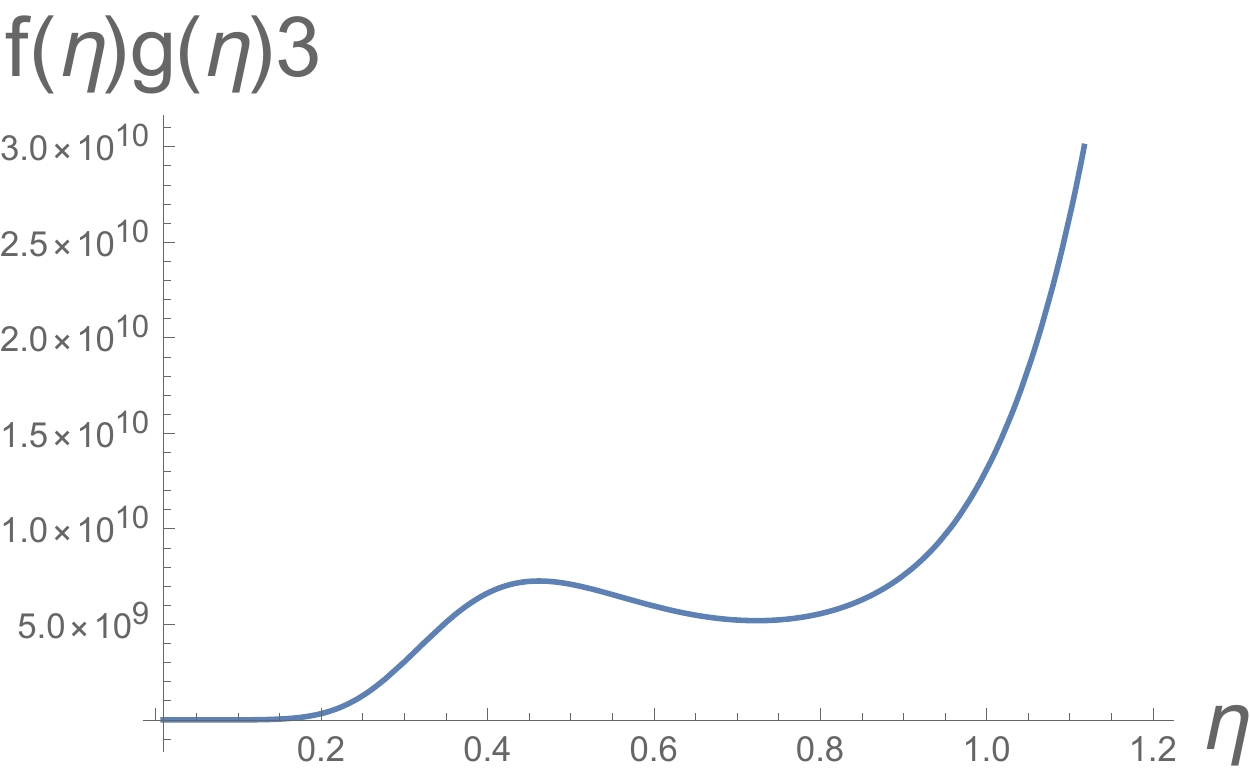}
\end{center}
\vspace{-5mm}
\caption{Plot of the metric function $f(\eta) g(\eta)^3$
as a function of $\eta$ in units of $L$.
}
\label{fig:fg3}
\end{figure}

\begin{figure}
\begin{center}
\includegraphics[width=7.5cm]{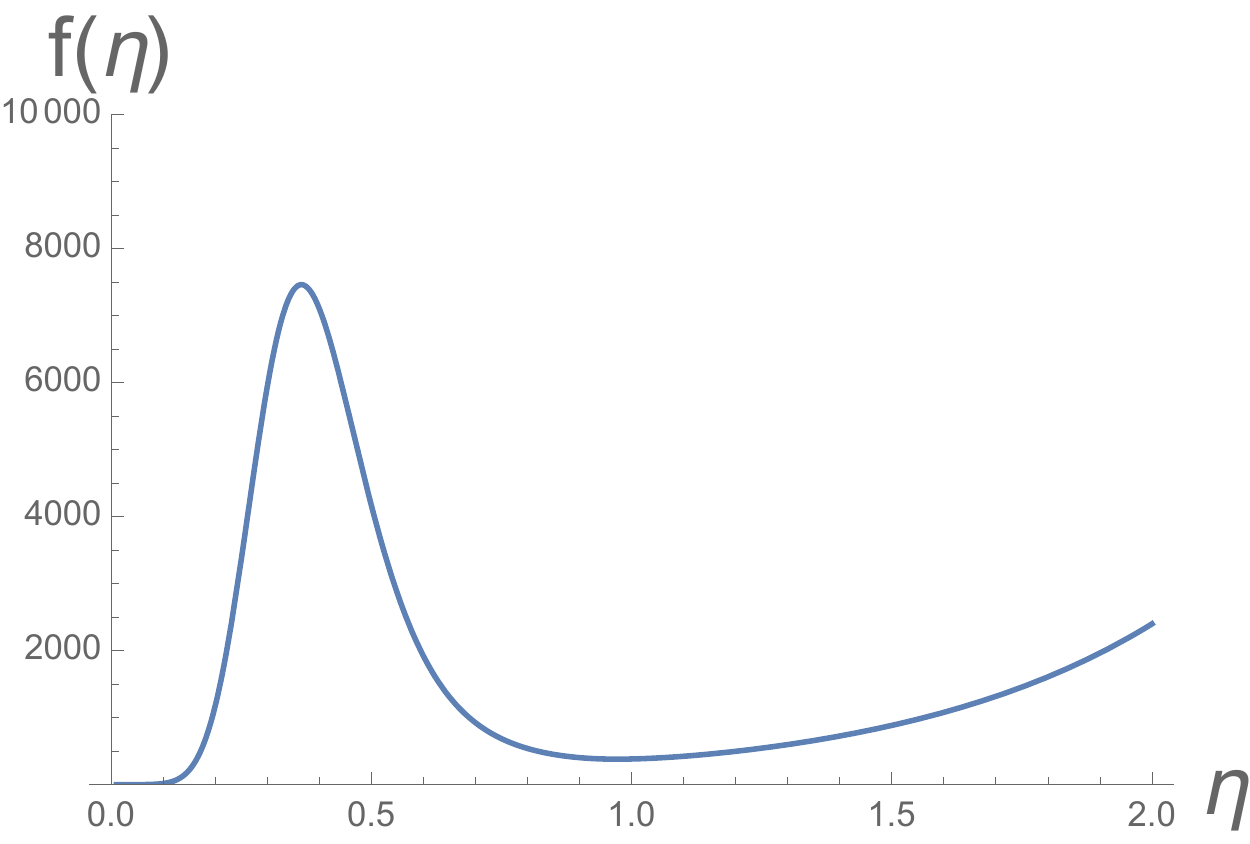}
\end{center}
\vspace{-5mm}
\caption{Plot of the metric function $f(\eta)$
as a function of $\eta$ in units of $L$.
}
\label{fig:f}
\end{figure}

For the choice (i), 
Fig.~\ref{fig:fg3} is the numerically obtained metric function $f(\eta) g(\eta)^3$, and Fig.~\ref{fig:f}
shows $f(\eta)$. The obtained $g(\eta)$ is 
\begin{align}
g(\eta) = c_5 e^{2\eta}
\end{align}
with $c_5=43.92$.
Another choice (ii) provides quite a similar plot for these functions.

%

%

Let us discuss a physical implication of the obtained metric. As clearly seen in Fig.~\ref{fig:f},
the temporal component of the metric has a peculiar feature: It has a big wall (bump) before it reaches
the black hole horizon $\eta=0$. This is different from the standard black hole metric. 
The AdS Schwartzschild black hole metric is $f(\eta)\propto \sinh^2[2\eta]/\cosh[2\eta]$,
which monotonically decreases when $\eta$ approaches the horizon $\eta=0$, so there is 
no wall. On the other hand, our $f(\eta)$ has a big wall.

The existence of the wall around $\eta = 0.4$ is due to the fact that the trained $h(\eta)$ is negative
around there. Integrating $h(\eta)$ to obtain the volume form $f(\eta)g(\eta)^3$ generates the wall.
The physical interpretation of the wall is striking: it resembles a {\it confining geometry}.
In a generic confining geometry in AdS/CFT, the curved spacetime ends at an IR wall.
The radial location of the wall corresponds to the energy scale of the confinement.
In our case, the deep learning automatically finds out that there exists such a confining
behavior, just from the data.

In fact, the output data of our neural network is arranged such that the geometry ends
with the black hole horizon. So, the emergence of the IR wall of the confining geometry
in the neural network is counter-intuitive. Nevertheless, the trained metric acquires the
wall to explain the data of the chiral condensate as a function of the quark mass.

Remember that in AdS/CFT, the boundary QFT is at the strong coupling limit, so the thermal
phase transition is at the first order. It means that there is no sense of a metric ``between"
the confining geometry and the black hole geometry. However, the realistic QCD is 
not at the strong coupling limit, and the thermal phase transition is a cross-over. We used the lattice data
at $T=207(5)$ [MeV] which is near the thermal phase transition, so the data is expected
to capture features of both the confining and the black hole geometries.
And indeed the trained metric acquires the features.

It is quite interesting that the metric trained in the deep learning automatically learns
the physical features of the phases. The wall behavior has not been studied in
the context of holographic QCD, to our knowledge, so it is a novel feature. This is the powerful point of
our deep learning holographic QCD modeling, which solves the inverse problem. 

%
\section{Quark confinement.}
\label{sec:5}

One of the most popular observables of QCD is the Wilson loop, and
its calculation in holographic QCD models is established \cite{Maldacena:1998im,Rey:1998ik,Rey:1998bq}.
As a concrete example of the ``prediction" part of our holographic QCD modeling in Fig.~\ref{fig:model},
we shall compute a quark antiquark potential in Sec.~\ref{sec:wil}.
The result is intriguing: the potential has a linear confining part, in addition to a part corresponding 
to the Debye screening.
That is, although our model at the temperature higher than
the critical temperature has a vanishing chiral condensate at $m_q=0$, the Wilson loop exhibits
a confining part.
We study in Sec.~\ref{sec:5-2} this interesting relation between 
the quark confinement and the chiral symmetry breaking, and discuss the
holographic origin of the relation.

\subsection{Calculation of Wilson loop.}
\label{sec:wil}

The calculation is straightforward, and we describe here only the formulas and results.
The Nambu-Goto string which hangs from the boundary of the asymptotic AdS spacetime,
whose ends are separated by the distance $d$, reaches $\eta=\eta_0$ at its deepest
in the radial $\eta$ direction. The relation between $d$ and $\eta_0$ is
\begin{align}
d = 2\int_{\eta_0}^\infty
\frac{1}{\sqrt{g(\eta)}}\sqrt{\frac{f(\eta_0)g(\eta_0)}{f(\eta)g(\eta)-f(\eta_0)g(\eta_0)}}d\eta \, .
\label{d}
\end{align}
The quark antiquark potential energy $V(d)$ is
\begin{align}
&
2\pi \alpha' V = 2\int_{\eta_0}^\infty \!\!\!\!
\sqrt{f(\eta)}
\left(
\sqrt{\frac{f(\eta_0)g(\eta_0)}{f(\eta)g(\eta)-f(\eta_0)g(\eta_0)}}
-1\right)d\eta 
\nonumber \\
&
\hspace{15mm} 
-2\int^{\eta_0}_0 \!\!\!\!
\sqrt{f(\eta)}d\eta\, .
\label{V}
\end{align}
Note that if $V(d)$ is positive, we should instead take $V(d)=0$ which corresponds to 
separated two quarks (two parallel Nambu-Goto strings stuck to the black hole horizon).
Eliminating $\eta_0$ in \eqref{d} and \eqref{V} provides the function $V(d)$.

\begin{figure}
\begin{center}
\includegraphics[width=7.5cm]{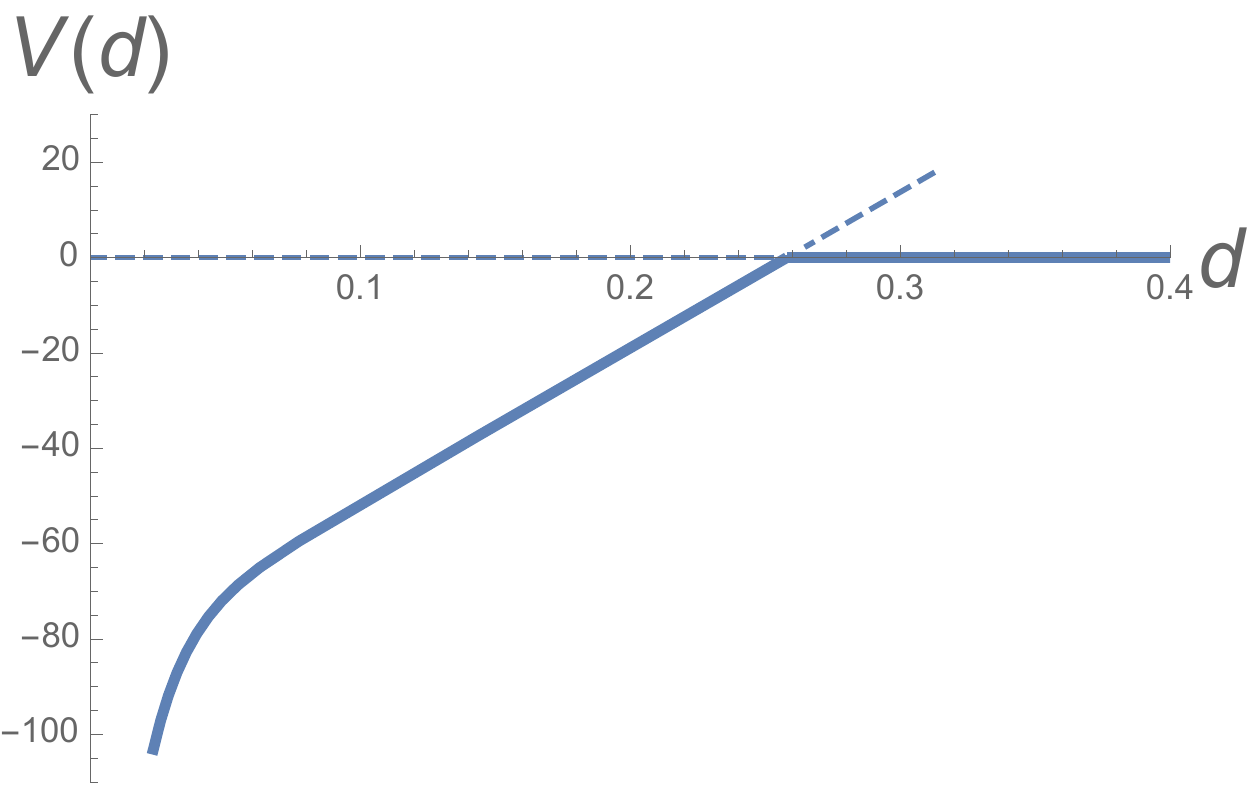}
\end{center}
\vspace{-5mm}
\caption{
Quark antiquark potential $V(d)$, as a function of the interquark distance $d$,
calculated holographically 
with the generated emergent metric using the case (i) ($g(\eta)\propto \exp (2\eta/L)$).
The distance $d$ is measured in units of
the AdS radius $L=0.6828$ [fm].
}
\label{fig:wilson}
\end{figure}

\begin{figure}
\begin{center}
\includegraphics[width=7.5cm]{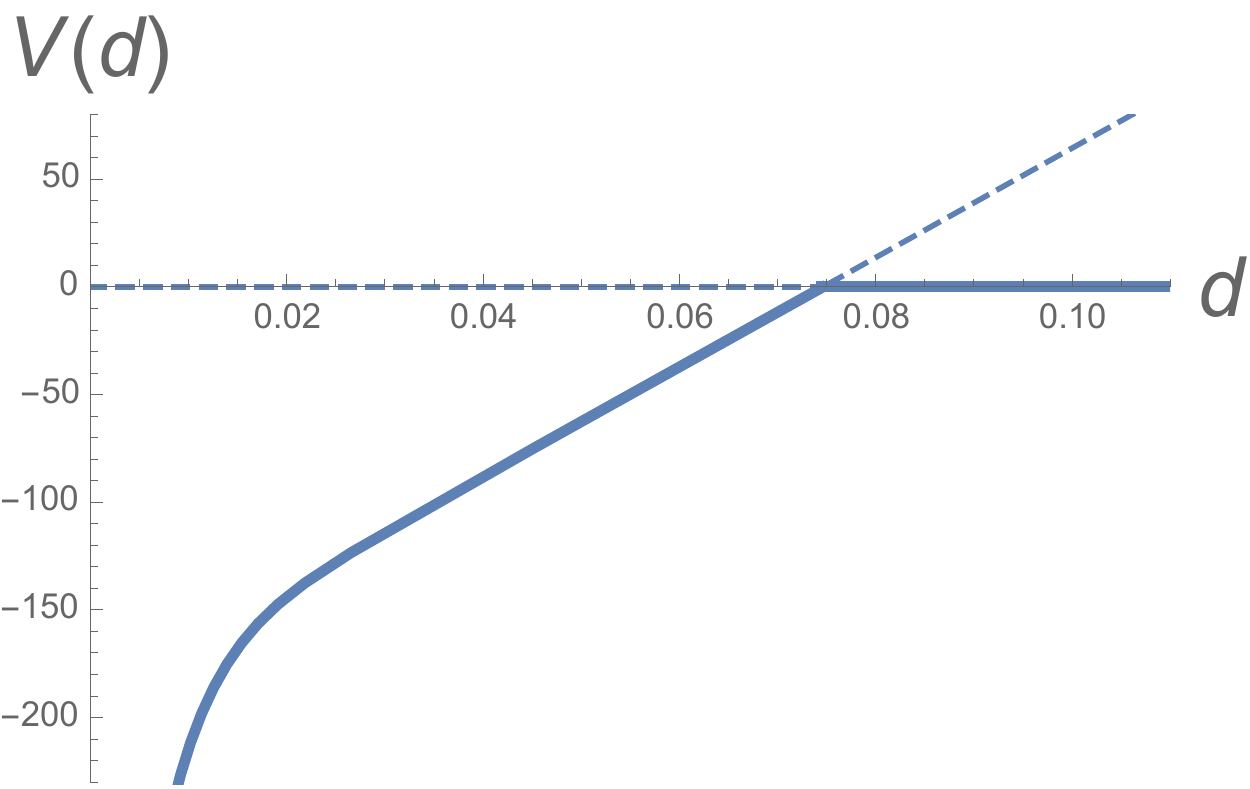}
\end{center}
\vspace{-5mm}
\caption{
Quark antiquark potential $V(d)$, using the case (ii) ($g(\eta)\propto \cosh (2\eta/L)$).
The shape is quite similar to the plot of Fig.~\ref{fig:wilson}, 
and the difference only appears in the overall scaling of the horizontal axis.}
\label{fig:wilson2}
\end{figure}

Using the trained metric determined in the previous section, the explicit form of $V(d)$
is calculated. The result for the case (i) is shown in Fig.~\ref{fig:wilson},
and the case (ii) is in Fig.~\ref{fig:wilson2}. (Since the overall
string tension $1/(2\pi\alpha')$ cannot be fixed in the holographic model, in the figures the
vertical axis is $2\pi \alpha' V(d)$.)

%
%

%
%

The important novel feature of the calculated quark antiquark potential 
is the co-existence of the linear potential and the Debye screening.
As clearly seen in the plot in Fig.~\ref{fig:wilson} and Fig.~\ref{fig:wilson2},
in addition to the flat part at larger $\eta$ which means the Debye screening
of the color-charged quarks, there exists the region of the linear potential
in the middle range of $\eta$. This co-existence has not been seen in previous 
holographic QCD models because 
the phase transition between the confining and the deconfining
phases is the first order at the strong coupling limit.

\begin{figure}
\begin{center}
\includegraphics[width=8.5cm]{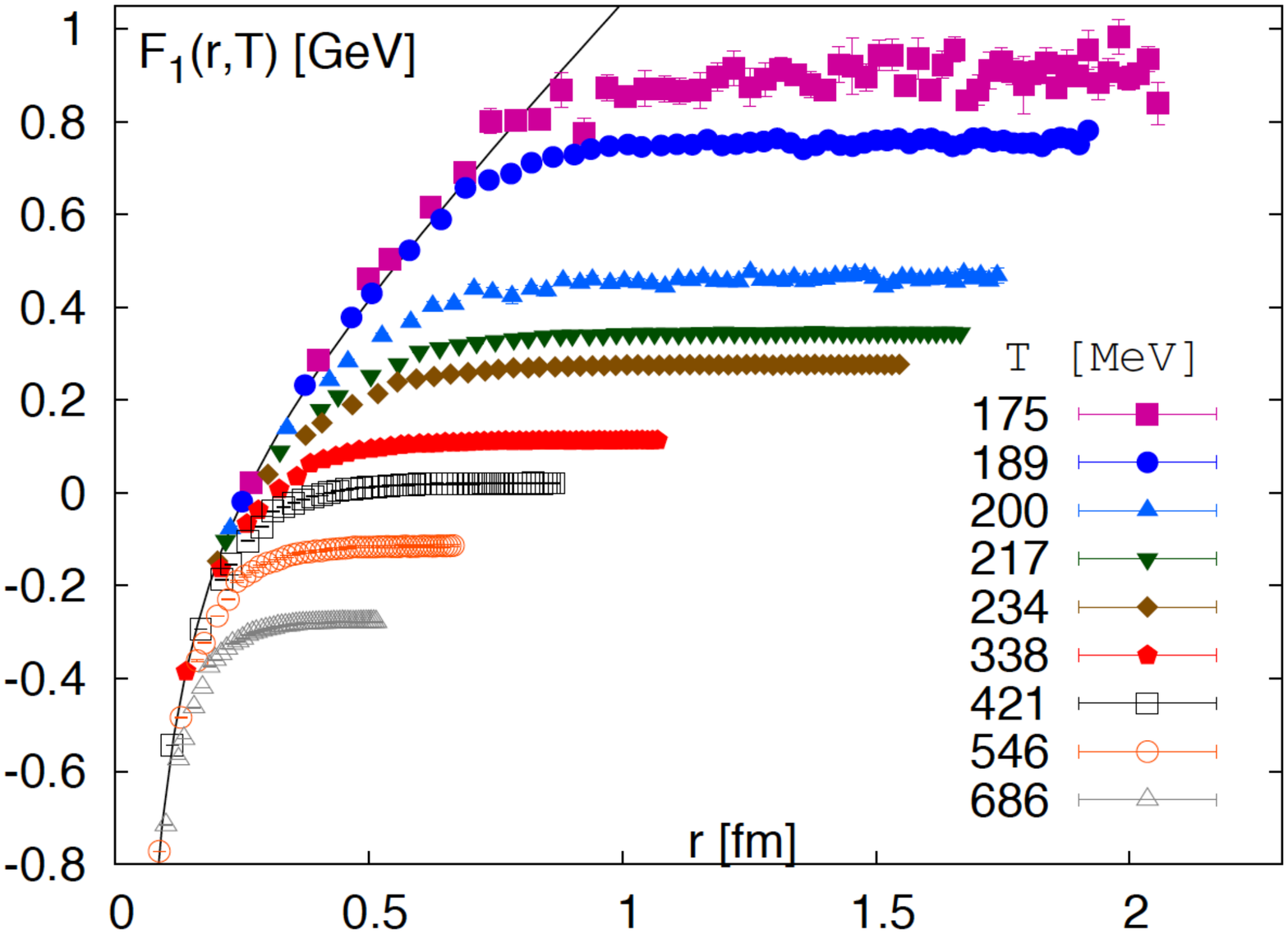}
\end{center}
\vspace{-5mm}
\caption{
A quark antiquark potential (singlet free energy) measured at finite temperature 
in lattice QCD. Figure taken from
Ref.~\cite{Petreczky:2010xg}. It shows coexistence of the Debye screening and
the linear confining potential.}
\label{fig:latticeqqbar}
\end{figure}

In fact, in the standard lattice QCD data of
the quark antiquark potential, one can find the coexistence of the Debye screening and
the linear confining behavior. For an example of such a lattice QCD data, see Fig.~\ref{fig:latticeqqbar}.
In this manner, the deep learning method can reproduce unexpectedly the important features
of physical observables (which are not used for the training of the network), from the emergent geometry.

%
%

\subsection{Quark confinement and chiral symmetry breaking.}
\label{sec:5-2}

It is often discussed in literature whether the quark confinement and the chiral symmetry breaking
are independent of each other or not 
(see for example Refs.~\cite{Suganuma:1993ps,Miyamura:1995xn,Woloshyn:1994rv,Fukushima:2002ew,Hatta:2003ga,Gattringer:2006ci,Bilgici:2008qy,Synatschke:2008yt,Lang:2011vw,Gongyo:2012vx,Glozman:2012fj,Doi:2014zea,Suganuma:2017syi,Suganuma:2016lnt} for various study on this question). 
In realistic QCD, the thermal deconfinement phase transition occurs at a temperature close to 
that of the chiral transition, that brings about the natural question.
Here, in our holographic model, the deep learning found a novel emergent metric which possesses
both the features of the confinement and the deconfinement phases, and the metric itself has 
been determined by the lattice data of the chiral condensate. 
So our holographic model is a good arena for discussions about the relation between the confinement 
and the chiral symmetry breaking.

Since our lattice data used to feed the neural network is the one measured at $T=207(5)$[MeV] which
is above the critical temperature, the chiral condensate at $m_q=0$ vanishes. So there is no spontaneous breaking of
the chiral symmetry. Nevertheless, the Wilson loop, or the quark-antiquark potential of probe quarks,
exhibits a linear confining part, as seen in Figs.~\ref{fig:wilson} and \ref{fig:wilson2}.
This leads to a conclusion that the chiral symmetry breaking is not directly related to the quark confinement,
in our simplest holographic model.

Even though our result of the calculation of the Wilson loop suggests that there is no
direct relation to the chiral symmetry breaking, if we look more carefully how the Wilson loop and
the chiral condensate are calculated in the holographic model, we find an intimate relation between
them. 

As we have seen in Sec.~\ref{sec:3}, since $h$ is the radial derivative of the metric, 
the negative $h$ means that there exists a wall in $f(\eta)$.
This wall produces the linear potential in the quark antiquark potential, as demonstrated explicitly 
above. In fact, an infinitely high wall is the IR confining wall which is often used in any holographic
QCD model of confinement. Our wall is of a finite height, and leads to the partially linear behavior 
of the quark antiquark potential. 

Since the confinement is attributed to the wall in the metric, let us study the chiral condensate from the view point 
of the metric. The equation for the bulk scalar field \eqref{sceq}, at the linear level in $\phi$, is
\begin{align}
\left[\frac{\omega^2}{f}+\partial^2_\eta  + h(\eta) \partial_\eta  +3\right] \phi = 0 \, .
\end{align}
Here we introduced a time-dependence $\phi \propto \exp(-i\omega t)$ so that
we can find the energy dependence of the modes explicitly.
Redefining the scalar field as
\begin{align}
\phi (\eta,t)= I(\eta) \tilde{\phi}(\eta,t)
\end{align}
with $\partial_\eta I/I=(-3/4)\partial_\eta g/g$, we find that the equation above is rewritten as
\begin{align}
\left[\sqrt{f}\partial_\eta \sqrt{f}\partial_\eta + \omega^2-V(\eta)\right] \tilde{\phi} = 0 \, ,
\end{align}
with
\begin{align}
V(\eta) \equiv -3f -f\frac{\partial_\eta^2I}{I}-hf\frac{\partial_\eta I}{I}\, .
\end{align}
Introducing a new radial coordinate $y = y(\eta)$ which satisfies
\begin{align}
\sqrt{f} \frac{d}{d\eta} = \frac{d}{dy}, 
\end{align}
the equation above can be recast to
\begin{align}
\left[-\frac{d^2}{dy^2} + V(\eta(y)) \right] \tilde{\phi} = \omega^2 \tilde{\phi} \, ,
\end{align}
which is of the form of a Shr\"odinger equation, with the energy eigenvalue $\omega^2$.

\begin{figure}
\begin{center}
\includegraphics[width=7.5cm]{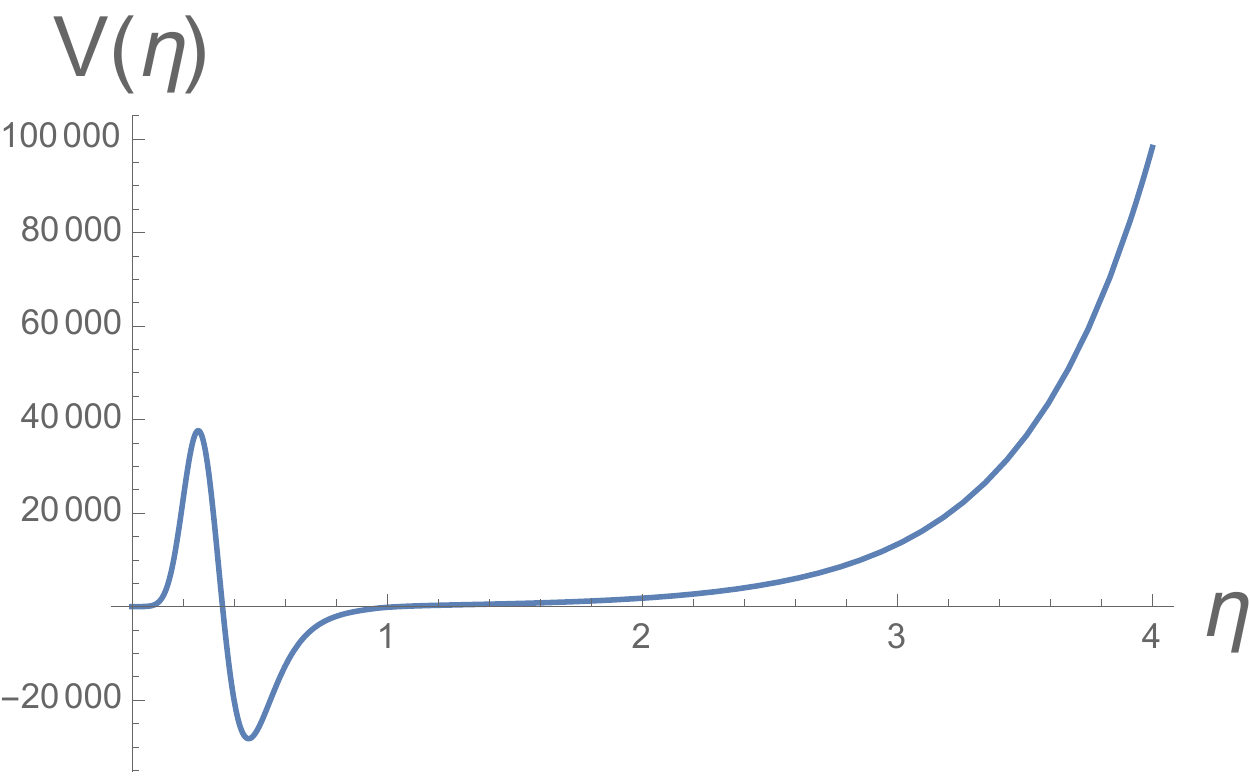}
\end{center}
\vspace{-5mm}
\caption{The potential $V(\eta)$ felt by the fluctuation of the bulk scalar field $\phi(\eta)$.}
\label{fig:V}
\end{figure}

For our previous choice $g(\eta)=c_5 e^{2\eta}$, the calculation of the Schr\"dinger potential
is simplified and we find 
\begin{align}
V(\eta) = \frac{3}{4}\left( \partial_\eta f-f\right) \, .
\end{align}
See Fig.~\ref{fig:V} for a plot of the potential $V(\eta)$. Because of the wall in $f(\eta)$,
resulting potential $V(\eta)$ has a deep valley which is separated away from the black hole horizon 
$(\eta=0)$ by a wall.
%
%
In general, a normalizable mode localized at the valley is expected. The chiral condensate means
the existence of such a mode with $\omega=0$. 
Our metric does not have such a mode, since our chiral condensate vanishes.
However, the potential $V(\eta)$ has a deep valley due to the
confining wall of $f(\eta)$, and it could have generated a chiral condensate.
This means that the wall in $f(\eta)$ can generate both the confinement and the chiral condensate,
and whether they actually develop in physical quantities depend on more details.

Another supportive argument for a relation between the chiral condensate and the wall in the metric is as follows.
For a generic operator in AdS to be condensed,
the bulk mass has to violate the Breitenlohner-Freedman (BF) bound 
\cite{Breitenlohner:1982bm,Breitenlohner:1982jf}. The BF bound signals a tachyonic instability of
any bulk field, which for the scalar field case is given by 
\begin{align}
m^2 +\frac{d^2}{4L^2}\geq 0 \, .
\label{BFb}
\end{align}
This bound is equivalent to the reality condition of the conformal dimension of the operator ${\cal O}$,
as seen from the AdS/CFT formula \eqref{DeltaO}. 
Since our metric function $h$ measures effectively the combination $d/L$ (since at a pure AdS spacetime 
we have $h=d/L$), and $d^2/L^2 \simeq h^2$ appears in the inequality above, 
we notice that a smaller $h(\eta)$ compared to its value at the spatial infinity $h(\eta=\infty)=d/L$
can generate the tachyonic instability and resultantly the chiral condensate $\langle \bar{q}q\rangle$.\footnote{
The condition \eqref{BFb} is only for pure AdS case while here we use it for the interior of the bulk.
Precisely speaking, the argument here is not quantitative, and further detailed calculation of
the tachyonic instability is necessary.}
In fact, the emergent geometry which the deep learning
obtained (Fig.~\ref{fig:metric}) has a region in which the value of $h$ is smaller than that
of the asymptotic value $h=4$ in units of $L=1$. It 
becomes even negative, which results in the wall structure of $f(\eta)$.
So, this negative region could have been an origin generating the chiral condensate. 
In other words, possible chiral symmetry breaking is due to the wall in the metric $f$.

We conclude that the wall structure in the metric could be a common 
origin of both the chiral symmetry breaking
and the quark confinement. 
These two phenomena could be generated from the simple wall structure in the bulk, 
in the simplest holographic setup.
However, as our calculated results of the holographic QCD model shows that
the quark antiquark potential has a confining part although the input data
of the chiral condensate at $m_q=0$ vanishes. Therefore, the actual relation between
the chiral symmetry breaking and the quark confinement is more subtle.\footnote{Of course, 
our analysis is at a single
temperature, and it is desirable to perform more analyses with different values of the temperature
to find more quantitative relation between the chiral symmetry breaking and the quark confinement.}

\section{Summary and discussion.}
\label{sec:6}

In this paper, we used a map \cite{Hashimoto:2018ftp} between the deep learning and  the AdS/CFT
to obtain explicitly a holographic QCD model. The bulk metric function was determined as neural network 
weights, and as the input data for the training of the neural network, 
we chose the lattice QCD data \cite{Unger} of the chiral condensate $\langle \bar{q} q\rangle$ 
as a function of the quark mass $m_q$ of QCD.
The emergent metric function turned out to have a unique structure: it has a wall and the horizon.
We calculated the quark antiquark potential (Wilson loop) using the bulk metric, and found that
the potential has both the linear potential part and the Debye screening part,
which coincides qualitatively with the quark antiquark potential obtained in lattice QCD.

It should be emphasized that the emergent metric has both the features of 
the confinement phase and the deconfinement phase, at the same time.
As mentioned in Sec.~\ref{sec:1}, normally in holographic QCD models
the two phases are completely separate, and there exists a first order phase transition
separating the two phases. In our case, we trained the neural network using 
the lattice QCD data, and the realistic QCD has a cross-over phase transition.
We attribute to the training data the reason why the emergent metric gained 
automatically both of the phases.

With this new feature of the metric, we could discuss that both the quark confinement
and the chiral symmetry breaking are possibly generated by the wall structure of the metric in the bulk.
However, our model has the vanishing chiral condensate at $m_q=0$ while
the quark antiquark potential has a confining part, thus there exists a discrepancy between
the quark confinement and the chiral symmetry breaking.

In this manner, our approach explores a wider class of holographic QCD models.
Using the realistic QCD data not only makes the holographic model more precise
(as it is ensured to reproduce the training data) but also provides unexpectedly
novel parameters of the model (which, in our case, the wall-shaped metric 
of the bulk).

If we look at the calculated Wilson loop (Fig.~\ref{fig:wilson}) more carefully,
the string-breaking distance $d$ is ${\cal O}(0.1)$[fm], while the lattice QCD result
in Fig.~\ref{fig:latticeqqbar} has a larger string-breaking distance. The reason of this
discrepancy would be allocated to the limitation of the holographic model Lagrangian
we employed. 
Note that our model is understood as a probe limit of the meson sector. 
Since our training data has many sets of pairs $(m_q, \langle\bar{q}q\rangle)$ to determine
a single metric, the obtained metric does not take into account any back reaction from
the bulk scalar field. 
Although incorporating a back reaction would make the neural network complicated,
it may lead to more unexpected features of the geometry. We leave it as a future work.

QCD is a QFT which has been most widely and deeply studied,
thus there exists tremendous amount of data on the physical observables of QCD.
Therefore, to explore the mystery of the AdS/CFT correspondence and the emergent  
bulk spacetime using data science,  QCD is the most suitable playground.
Spatial structure of the emergent geometry, and possible gravitational degrees of 
freedom emergent in the bulk, are of interest, and we expect that deep learning method
will provide us with more intuitions on those interesting issues.

\vspace{5mm}
\begin{acknowledgments}
We would like to thank W.~Unger for providing us with the lattice data \cite{Unger}.
K.~H.~would like to thank Elias Kiritsis for valuable discussions.
A.~Tanaka would like to thank Lei Wang for notifying us the nice paper.
The work of K.~H.~was supported 
in part by JSPS KAKENHI Grants No.~JP15H03658, 
No.~JP15K13483, and No.~JP17H06462. 
S.~S.~is supported in part by the Grant-in-Aid for JSPS Research Fellow, Grant No.~JP16J01004.
The work of A.~Tanaka was supported by the RIKEN Center for Advanced Intelligence Project and JSPS KAKENHI Grants No.~JP18K13548.
A.~Tomiya was fully supported by Heng-Tong Ding.
The work of A.~Toimya was supported in part by NSFC under grant no. 11535012
and the RIKEN Special Postdoctoral Researcher program.
\end{acknowledgments}

\vspace{5mm}

\appendix
\section{Details about our training code.}
\label{app:1}

For generating the training data $\{ (\bar{x}^{(1)} , \bar{y}) \}$ where $\bar{y}=0$ ($\bar{y}=1$) 
corresponds to the positive (negative) data, we follow the steps described below.
First, we plot the data of the right panel of Table \ref{latticedata} $(m_q, \langle\bar{q}q\rangle)$ 
to a 2-dimensional scattered plot and fit it by a polynomial with respect to $m_q$ up to the 5-th order, and call it $f(m_q)$.
By using this $f(m_q)$, we prepare the training data $\{ (\bar{X}^{(1)}, \bar{y} ) \}$ as follows:
\begin{enumerate}
\item
Randomly choose $m_q \in [0 , 0.022]$, $\langle \bar{q}q\rangle \in [0, 0.11]$.
\item
Convert 
$(m_q, \langle \bar{q}q\rangle)$
 to $\phi(\eta_{\rm ini})$ and $\pi(\eta_{\rm ini})$ by \eqref{phiini} and \eqref{piini},
and regard them as the input $\bar{x}^{(1)}$.
\item
Define the answer signal : 
$$
\bar{y}
=
\left\{ \begin{array}{ll}
0 & \text{if } \langle \bar{q}q\rangle \in [f(m_q) - \text{noise},  f(m_q)+\text{noise}] \\
1 & \text{otherwise}\\
\end{array} \right.
$$
where the noise is sampled from a gaussian with the average 0 and the standard deviation 0.004.
\end{enumerate}
The total training data $D$ consists of:
\begin{align}
D
&=
\Big(
\text{$10^4$ positive data}
\Big)
\oplus
\Big(
\text{$10^4$ negative data}
\Big),
\nonumber
\end{align}
where
\begin{align}
\left\{ \begin{array}{ll}
\text{positive data} = \{ (\bar{x}^{(1)}, \bar{y}=0) \}  \\
\text{negatve data} = \{ (\bar{x}^{(1)}, \bar{y}=1) \} 
\end{array} \right.
.
\notag
\end{align}
To compare $\bar{y}$ and the neural network output $y$, at the final layer
we calculate $F \equiv \pi(\eta_\text{fin})$ (which is the r.h.s.~of \eqref{Ff} in the limit $\eta_\text{fin}\to 0$), 
and then we define $y \equiv t(F)$ with
\begin{align}
t(F)
\equiv
\frac{\tanh\Big(100 (F - 0.1) \Big)}{2}
-
\frac{\tanh\Big(100 (F + 0.1) \Big)}{2}
+1.
\end{align}
This function measures how $F=\pi(\eta_\text{fin})$ is small: roughly speaking, $t(F)\approx 0$ for $|F|<0.1$ and $t(F)\approx 1$ for $|F|>0.1$. 

As the training, we repeat the following training iteration:
\begin{enumerate}
\item{Randomly divide the training data to a direct sum of size 100 mini data: $D =(\text{mini data})_1 \oplus (\text{mini data})_2 \oplus \dots \oplus (\text{mini data})_{200}  $.}
\item{Calculate loss \eqref{loss} and update $h(\eta^{(n)})$ by Adam optimizer \cite{kingma2014adam} for each mini data.}
\end{enumerate}
Hyper-parameters of the Adam optimizer are taken as follows:
$\alpha=0.002, \beta_1=0.9, \beta_2=0.999, \epsilon=10^{-8}$.
When the loss \eqref{loss} becomes less than 0.08, we stop the iteration 1 and 2.


\section{Normalization of the two-point function}
\label{app:2}

When we map the boundary data to the boundary behavior of the bulk scalar, we need to take care of the normalization of the operator.   
Suppose that the bulk scalar $\phi$ is dual to the CFT operator $\mathcal{O}$ with dimension $\Delta$. 
If the real scalar $\phi$ has the canonical kinetic term and the boundary behavior is as follows\footnote{Note that we set $L=1$ in this appendix.} 
\begin{align}
\phi  \approx
e^{-(d-\Delta)\eta}  b J + e^{-\Delta\eta}\frac{\braket{\mathcal{O}}}{b(2 \Delta-d)}\, ,    
\end{align}
the standard holographic computation \cite{Klebanov:1999tb} says that the two point function of $\mathcal{O}$ results in 
\begin{align}
\braket{\mathcal{O}(x)\mathcal{O}(0)}=b^2 \frac{(2\Delta-d)\pi^{-\frac{d}{2}}\Gamma(\Delta)}{\Gamma(\Delta-\frac{d}{2})}\frac{1}{|x|^{2\Delta}}\, .
\label{eq_b2}
\end{align}
Here, the constant $b$ can be used to tune the normalization without changing the coefficient of the source term  $J \mathcal{O}$ (see  Ref.~\cite{Cherman:2008eh}).

We now argue  that this $b$ should be set as $b=\frac{\sqrt{N
	_c}}{2\pi}$ for the case of $\mathcal{O}=\bar{q}q$ where the Dirac fermion $q$ has the canonical normalization. 
The canonical dimension of $\bar{q}q$ in $d=4$ is $\Delta=3$. Thus, \eqref{eq_b2} becomes 
\begin{align}
\braket{\mathcal{O}(x)\mathcal{O}(0)}=b^2 \frac{4  }{\pi^2 |x|^{6}}
\label{eq_O2_norm}
\end{align}  

On the other hand, the canonical Euclidean Lagrangian of free Dirac fermion  is 
\begin{align}
\mathcal{L} = \bar{q}_{a} (\slashed{\partial} +m_q)q_{a} ,\quad (a=1,\cdots, N_c) \, .
\label{free_fermion}
\end{align}
The Wick contraction gives  
\begin{align}
\contraction{}{q}{_a(x)}{\bar{q}} q_a(x) \bar{q}_b(0)= \delta_{ab} \int \frac{d^4 p}{(2\pi)^4} e^{i p x} \frac{-i\slashed{p}+m}{p^2+m^2}\, .
\end{align}
Using this propagator, 
we can compute the two-point function of the composite operator $\mathcal{O}= \bar{q}q$ at the Gaussian fixed point ($m_q=0$). 
It is given by 
\begin{align}
&\braket{\mathcal{O}(x)\mathcal{O}(0)}\nonumber\\
&=-N_c \int \frac{d^4 q}{(2\pi)^4}
e^{i q x}
\int \frac{d^4 k}{(2\pi)^4}\tr \left[\frac{-i\slashed{k}}{k^2}\cdot \frac{i(\slashed{q}-\slashed{k})}{(q-k)^2}\right] \, .
\end{align}
We evaluate the $k$-integral by using the dimensional regularization ($d=4-\epsilon$): 
\begin{align}
I(q)=\int \frac{d^d k}{(2\pi)^d}\tr \left[\frac{-i\slashed{k}}{k^2}\cdot \frac{i(\slashed{q}-\slashed{k})}{(q-k)^2}\right] \, .
\end{align}
Using the Feynman integral formula, the loop integral becomes 
\begin{align}
I(q)
&=-d  \int^1_0 dx  \int \frac{d^d \ell}{(2\pi)^d} \frac{\ell^2-x(1-x)q^2}{[\ell^2 +x(1-x)q^2]^2}\, .
\end{align}
We now use the formula 
\begin{align}
&\int \frac{d^d \ell}{(2\pi)^d}  \frac{1}{[\ell^2+ \Delta]^2}= \frac{\Gamma(2-\frac{d}{2})}{(4\pi)^{\frac{d}{2}}} \Delta^{\frac{d}{2}-2}\, ,\\
&\int \frac{d^d \ell}{(2\pi)^d}  \frac{\ell^2}{[\ell^2+ \Delta]^2}= \frac{d\,\Gamma(1-\frac{d}{2})}{2(4\pi)^{\frac{d}{2}}} \Delta^{\frac{d}{2}-1}\, .
\end{align}
Then the loop integral is given by 
\begin{align}
I(q)
&=
-\frac{2d(d-1)\Gamma(2-\frac{d}{2})\Gamma(\frac{d}{2})^2}{(2-d)(4\pi)^{\frac{d}{2}}\Gamma(d)} |q|^{d-2}\, .
\end{align}
By expanding it around $\epsilon=0$ where $d=4-\epsilon$, we can find that $I(q)$ has a finite term 
\begin{align}
-\frac{1}{8 \pi^{2}} q^2 \log q^2\, .
\end{align} 
Therefore, we have
\begin{align}
\braket{\mathcal{O}(x)\mathcal{O}(0)}= \frac{N_c}{8\pi^2} \int \frac{d^4q}{(2\pi)^4} e^{-i q x} \left[q^2 \log q^2+\cdots \right] \, .
\label{qqbar_normalization}
\end{align}

We compare this result with \eqref{eq_O2_norm}. By a similar dimensional regularization approach,  one can find that the Fourier transformation of \eqref{eq_O2_norm}, 
\begin{align}
\int d^d x e^{i q x} \braket{\mathcal{O}(x)\mathcal{O}(0)}=b^2 \int d^d x e^{i q x} \frac{4  }{\pi^2 |x|^{6}}
\end{align}
has a term containing $q^2 \log q^2$. It is given by 
\begin{align}
\frac{b^2}{2}q^2 \log q^2\, .
\end{align} 
Comparing it to \eqref{qqbar_normalization}, we conclude 
\begin{align}
b=\frac{\sqrt{N_c}}{2\pi}\, .
\end{align}


\end{document}